\documentclass[prb,amsmath,amssymb,superscriptaddress]{revtex4}

\usepackage{graphicx}
\usepackage{color}
\usepackage{amsmath}
\usepackage{bm}

\begin{document}

\title{Surface superconductivity in rhombohedral graphite}

\author{N.~B.~Kopnin}

\affiliation{Low Temperature Laboratory, Aalto University, P.O. Box 15100, FI-00076 AALTO, Finland}

\affiliation{ L.~D.~Landau Institute for
Theoretical Physics, 117940 Moscow, Russia}

\author{T.~T.~Heikkil\"a}

\affiliation{Low Temperature Laboratory, Aalto University, P.O. Box 15100, FI-00076 AALTO, Finland}

\newcommand{\tmpnote}[1]%
   {\begingroup{\it (FIXME: #1)}\endgroup}
   \newcommand{\comment}[1]%
       {\marginpar{\tiny C: #1}}


\begin{abstract}
We show that rhombohedral graphite may support surface superconductivity with an unusual relation between the BCS coupling constant and the order parameter. This feature results from the properties of the states localized on the graphite surfaces. In a description including only the nearest neighbour coupling of the graphene layers, the surface states are topologically protected and have a flat band dispersion. We show that including higher order couplings destroys this flat band character and leads to a particle-hole symmetry breaking quadratic dispersion with a large effective mass. Employing this dispersion, we then show its effect on superconductivity and find two regimes of parameters, depending on the relation between the strength of the coupling constant and the details of the quadratic dispersion. For low coupling strengths, superconductivity is localized on the surfaces, but the order parameter is exponentially suppressed as in a conventional BCS superconductor, whereas for large coupling strengths we obtain surface superconductivity with a linear relation between the order parameter and the coupling constant. Our results may explain the recent findings of graphite superconductivity with a relatively high transition temperature.
\end{abstract}

\date{\today}

\maketitle

\section{Introduction}
\label{sec:introduction}

Superconductivity is a ubiquitous phenomenon in metals: according to a commonly held view, all metals become either superconducting or magnetic at low enough temperatures. However, the corresponding transition temperatures may be (so far) unobservably low. In conventional superconductors, such as Al, Hg, or Nb, the transition temperature depends exponentially on the inverse of the BCS coupling constant, and whereas the coupling constant itself may be fairly large (the relevant energy scale connected with it may be many times larger than the thermal energy at room temperature), the resulting transition temperature typically does not much exceed 10 K. This property is intrinsically related to the quadratic dependence of electrons' energy on momentum, which leads to a logarithmic divergence in the BCS self-consistency equation. With a higher-order dispersion around the Fermi energy, the relation between the magnitudes of the critical temperature and the coupling constant becomes stronger, boosting the superconductivity. 

The extreme case would be a completely dispersionless energy spectrum, the so called ``flat band''.
Fermionic systems with dispersionless branches of excitation
spectrum have quite unusual properties; nowadays they
attract lots of research interest. Flat bands were predicted in
many condensed matter systems, see for example
\cite{Khodel1990,NewClass,Shaginyan2010,Gulacsi2010}. In
some cases the flat bands are protected by topology in momentum
space; they emerge  on the surfaces of gapless topological
matter\cite{HeikkilaKopninVolovik10} such as surfaces of nodal
superconductors \cite{Ryu2002,SchnyderRyu2010,Brydon2011}, graphene edges
\cite{Ryu2002}, surfaces of multilayered graphene structures
\cite{GuineaCNPeres06,HeikkilaVolovik10-1,MakShanHeinz2010,Dora2011}, and
in the cores of quantized vortices in topological superfluids and
superconductors
\cite{HeikkilaKopninVolovik10,KopninSalomaa1991,Volovik2011}.
The singular density of states (DOS) associated with the
dispersionless spectrum was recently shown by us to essentially enhance the transition
temperature opening a new route to room-temperature
superconductivity. 

The problem is to find the metal with such a higher-order dispersion around the Fermi sea. Along with our collaborators, we have shown \cite{KopninHeikkilaVolovik2011,HeikkilaVolovik10-1} that within the nearest-neighbour approximation, rhombohedral graphite has topologically protected surface states with a flat band at the Fermi energy, and these surface states support high-temperature superconductivity where the superconducting order parameter is concentrated around the surfaces. Such a superconductor may also carry a large surface supercurrent with a critical value proportional to the large critical temperature. The corresponding critical temperature depends
linearly on the pairing interaction strength and can be thus
considerably higher than the usual exponentially small critical
temperature in the bulk. A flat band forms out of a low dispersive band that appears on the surface of a multilayered graphene structure with rhombohedral stacking with a large number of layers. 
Surface superconductivity is favorable already for a
system having $N\geq 3$ layers, where the normal-state spectrum
has a power-law dispersion $\xi_{p} \propto |{\mathbf p}|^N$ as a
function of the in-plane momentum ${\mathbf p}$. The DOS $\nu(\xi_{
p}) \propto \xi_{p}^{(2-N)/N}$ has a singularity at zero energy
which results in a drastic enhancement of the critical
temperature.

However, next-nearest neighbour hoppings which are present in real rhombohedral graphite can break the exact topological protection and, therefore, the flat-band mechanism of superconductivity at sufficiently low values of the coupling constant can be destroyed. Here we study the detailed effect of these higher-order interactions and show that though they indeed break the flat-band scenario for weak superconducting coupling, they provide another mechanism of surface superconductivity which is of the BCS type but has a much larger coupling constant than the usual superconductivity in bulk graphite. This large coupling constant comes from a large DOS associated with a heavy effective mass of surface quasiparticles that is clearly distinguishable on the background of the flat band which would exist without the higher-order interactions. Both these mechanisms favor the high-temperature superconductivity. Our results provide a criterion for the parameters needed to obtain the highest critical temperature. They
may be relevant in explaining the recent experimental findings \cite{Kopelevich01,Esquinazi08,Dusari10,Esquinazi12,Ballestar12} reporting the observation of even room-temperature superconductivity in doped graphite. 

Here we briefly outline the main results of our paper. More detailed calculations are described in the following sections.
In this work we study rhombohedral graphite by taking into account, besides the lowest-order interlayer hopping energy $\gamma_1$, also the higher-order hoppings $\gamma_3$ and $\gamma_4$. From these, only the latter breaks the topological protection of the flat band. We find that, even in the presence of all these interactions, rhombohedral graphite still has surface states. However, instead of a flat band of radius $p_{\rm FB}= \gamma_1/v_F$ they have a weak dispersion 
\begin{equation}
\epsilon_p=\alpha \left(\frac{p}{p_{\rm FB}}\right)^2-\mu \ ,
\end{equation}
within the region $p<p_{\rm FB}$. Here a small factor $\alpha=2 \gamma_1 \gamma_4 /\gamma_0$ arises from the higher-order interlayer hopping $\gamma_4$; $\gamma_0$ is the zero-order intralayer nearest-neighbour hopping energy, $v_F$ is the Fermi velocity in graphene, and $\mu$ is the chemical potential. These surface states are not symmetric with respect to the point $\epsilon=0$, more or less in the same way as the presence of the next-neighbour coupling breaks the electron-hole symmetry in graphene \cite{reviewCastroNeto09}. For $p>p_{\rm FB}$, the energy of the surface states deviates rapidly from $\mu$ (see Figs.~\ref{fig:3dspectrum} and \ref{fig:2dnormalspectrum}).  Qualitatively, the energy gap is localized at the surface and can be determined by a simplified self-consistency equation, which for $T=0$ and $\mu =0$ has the form 
\begin{equation}
1=\frac{g}{2\pi p_{\rm FB}^2}\int_0^{p_{\rm FB}} p dp \frac{1}{\sqrt{\Delta^2+\epsilon_p^2}} =\frac{g}{4 \pi \alpha}{\rm Arsinh}(\alpha/\Delta) \ ,
\label{eq:selfcons1}
\end{equation}
where $g=Wp_{\rm FB}^2/\hbar^2 d$ is a superconducting coupling energy proportional to the pairing interaction $W$, $d$ is the distance between the layers. Equation (\ref{eq:selfcons1}) yields
\begin{equation}
\Delta=\alpha \sinh^{-1}\left(4 \alpha \pi/g\right)\ .
\label{eq:simpleappr}
\end{equation}
We thus find that for $g < g_c \sim 4 \pi \gamma_1 \gamma_4/\gamma_0$, the gap $\Delta$ is exponentially small, similar to conventional BCS superconductors, whereas for larger coupling constant $\Delta$ tends to the flat-band result \cite{KopninHeikkilaVolovik2011} $\Delta\propto g$ which is linear in the coupling strength. These values of $\Delta$ yield critical temperatures $T_c$ of the order $\Delta/k_B$, the exact prefactor depending on which of the above regimes the coupling constant is.

The above considerations help to identify the regime of parameters where extremely high-temperature surface superconductivity might be found. In our estimates, we use the tight-binding parameters summarized in \cite{reviewCastroNeto09} (see also \cite{Dresselhaus02}), which gives $\gamma_0=3.2$ eV, $\gamma_1=0.39$ eV, $\gamma_3=0.315$ eV, and $\gamma_4=0.044$ eV. We disregard the next-nearest neighbour intralayer hopping proportional to $\gamma_2 =-0.02$ eV and even smaller interaction across two layers. In addition of its small relative magnitude, the next-nearest hopping $\gamma_2$ does not affect the conical dispersion of single graphene layer and thus is not expected to essentially modify our results. Note, however, that also widely different values of higher-order hopping constants are discussed in literature \cite{arovas08}, especially with a much higher $\gamma_4$ (for example \cite{McClure69} claims $\gamma_4=\gamma_3$). With the above values, the crossover between the two regimes takes place around $g_c \approx 0.1\dots 0.2 \gamma_1\approx$ 0.04\dots 0.08 eV (see Fig.~\ref{fig:selfconsdelta} below). Thus, for $g\ll g_c$, the gap and the critical temperature would be exponentially small whereas for $g>g_c$ it would be $T_c \sim (g/\gamma_1) \times 60$ K. This is much larger than the expected gap in the bulk for the same magnitude of the coupling.

This paper is organized as follows. In Sec.~\ref{sec:dispersion} we derive the dispersion of the surface states for rhombohedral graphite in the normal state by including the interlayer couplings $\gamma_1$, $\gamma_3$ and $\gamma_4$. In Sec.~\ref{sec:bdgequations} we derive the Bogoliubov-deGennes (BdG) equations for the superconducting state, while Sec.~\ref{sec:superconductivity} describes the surface superconductivity in rhombohedral graphite. Besides analyzing the crossover between the exponentially damped superconductivity and strong flat-band superconductivity, we review our earlier results on flat band superconductivity in the case of finite number of layers and on supercurrent carried by the surface states. In addition to this, we analyze the role of fluctuations around our mean-field solution and detail our previous prediction \cite{HeikkilaKopninVolovik10} of interface superconductivity at twinning layers of graphite by showing a few example cases. 
The paper is concluded by a short summary of our work and a comparison to the recent experiments on superconductivity in graphite.


\section{Electron dispersion in rhombohedral graphite}
\label{sec:dispersion}

\begin{figure}[h]
\centering
\includegraphics[width=0.8\linewidth]{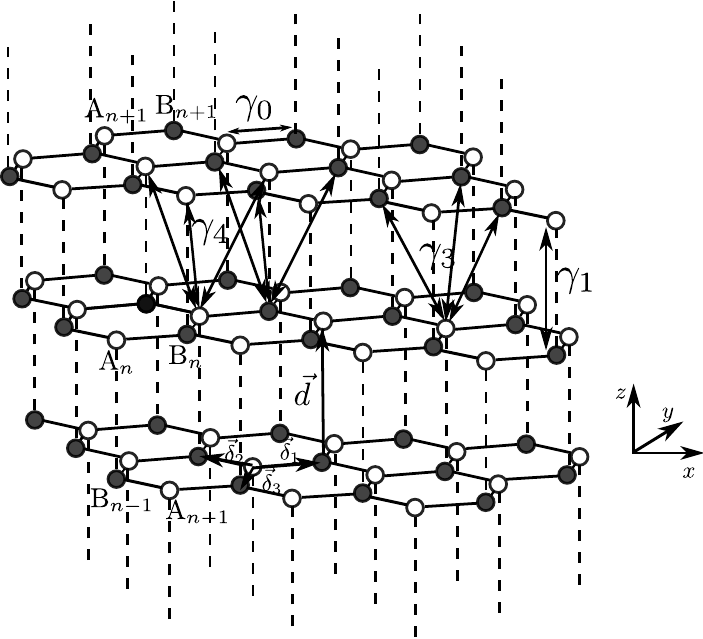}
\caption{(Color online) Rhombohedral graphite lattice, tight-binding parameters and the lattice vectors.}
\label{fig:rhggraphite}
\end{figure}

The rhombohedral graphite lattice and the tight-binding couplings are depicted in Fig.~\ref{fig:rhggraphite}. We denote the layers (starting from the bottom) by index $n$, the position of A atoms inside layer $n$ by $R_{i,n}$, the vectors from the A atoms to the nearest B atoms by ${\bm \delta}_j$ ($j=1,2,3$), and the vector between the layers by ${\mathbf d}$. The latter is strictly in the vertical direction, i.e., connects a B atom from layer $n$ to an A atom in layer $n+1$ (this is the convention leading to $\sigma_-$ coupling on the upper diagonal).

For this lattice structure we can write the Hamiltonian 
\begin{equation}
H=\sum_{l=0}^4 H^{(l)}\ , \label{eq:tbbilayer}
\end{equation}
where
\begin{eqnarray*}
H^{(0)}&=&-\gamma_0\sum_{n=1}^N\sum_{{\mathbf R}_{i,n}}\sum_{j=1,2,3}
\left[\psi_n^{A\dagger}({\mathbf R}_{i,n}) \psi_n^B({\mathbf R}_{i,n}+{\bm \delta}_j) + {\rm
h.c.}\right]\\
H^{(1)}&=& -\gamma_1 \sum_{n=1}^{N-1} \sum_{{\mathbf R}_{i,n}}  \left[\psi_n^{B\dagger}({\mathbf R}_{i,n}+{\bm \delta}_1) 
\psi_{n+1}^A({\mathbf R}_{i,n}+{\mathbf d}+{\bm \delta}_1) + {\rm h.c.}\right]\\
H^{(3)}&=& -\gamma_3 \sum_{n=1}^{N-1}\sum_{{\mathbf R}_{i,n}} \sum_{j=1}^3
\left[\psi_n^{A\dagger}({\mathbf R}_{i,n}) \psi_{n+1}^B({\mathbf R}_{i,n}+{\mathbf d} -{\bm \delta}_j)+{\rm h.c.}\right]\\
H^{(4)}&=& -\gamma_4 \sum_{n=1}^{N-1} \sum_{{\mathbf R}_{i,n}}  \sum_{j=1}^3 \left[\psi_n^{A\dagger}({\mathbf R}_{i,n}) 
 \psi_{n+1}^A({\mathbf R}_{i,n}+{\mathbf d}+{\bm \delta}_j) + {\rm h.c.}\right]\nonumber \\
 && -\gamma_4 \sum_{n=1}^{N-1} \sum_{{\mathbf R}_{i,n}} \sum_{j=1}^3 \left[\psi_n^{B\dagger}({\mathbf R}_{i,n}+{\bm \delta}_1) 
  \psi_{n+1}^B({\mathbf R}_{i,n}+{\mathbf d}+{\bm \delta}_1+{\bm \delta}_j)\right.\nonumber \\
 &&  +\left. {\rm h.c.}\right]\ .
\end{eqnarray*}
The sum over ${\mathbf R}_{i,n}$ goes over all A-atoms in layer $n$.  For the hopping constants we use the most recent data in graphite according to \cite{reviewCastroNeto09,Dresselhaus02}: $\gamma_0=3.2$ eV, $\gamma_1=0.39$ eV, $\gamma_3=0.315$ eV and $\gamma_4=0.044$ eV and neglect the intra-layer next-nearest neighbour hopping term $H^{(2)}$ proportional to $\gamma_2=- 0.02$ eV, as well as the even smaller hoppings across two layers.

\begin{figure}[t]
\centering
\includegraphics[width=0.8\linewidth]{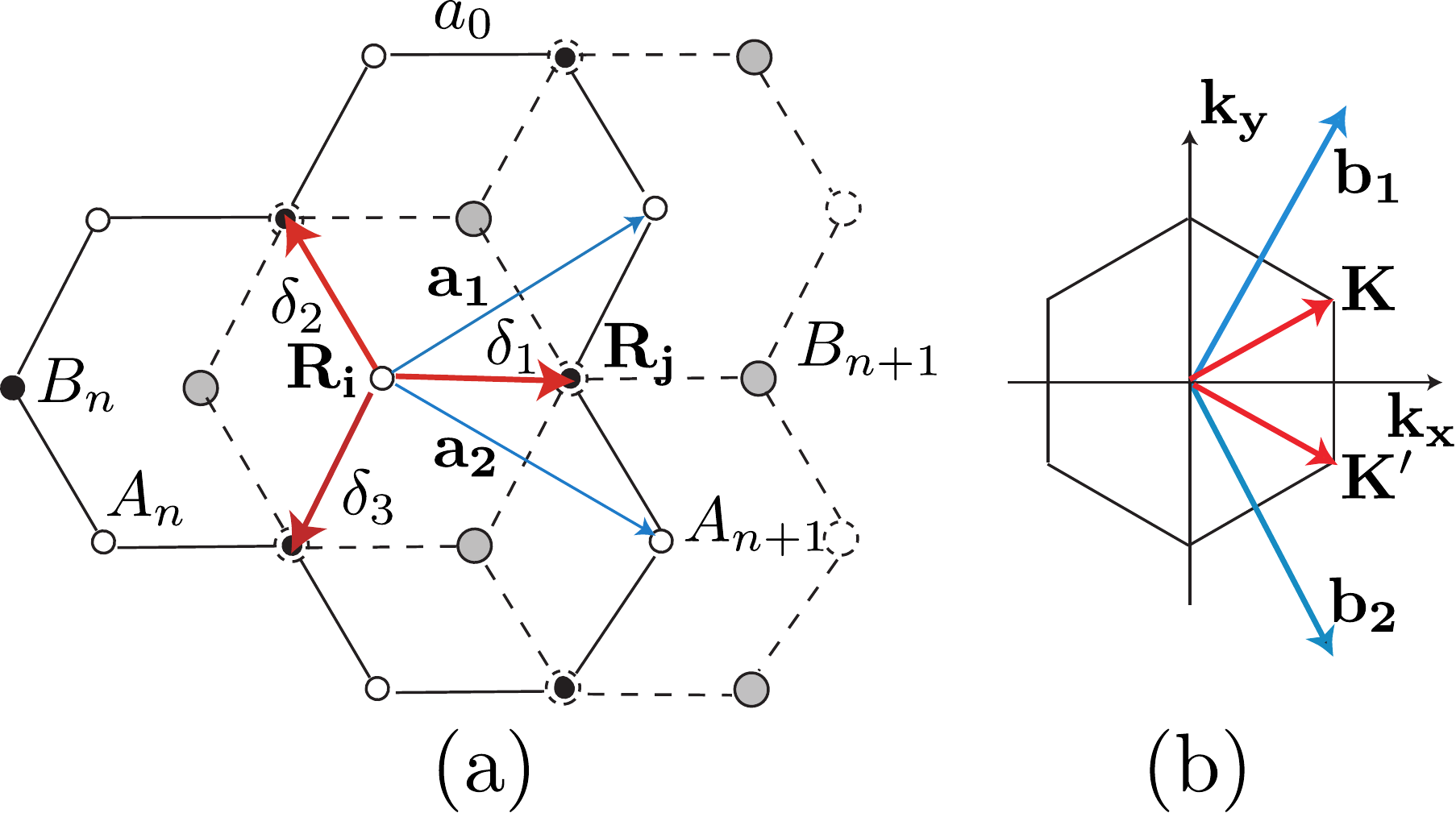}
\caption{(Color online) (a) Projection  of the unit cell onto the layer plane (from bottom to top). Layer $n$ is shown by solid circles and lines, while the next layer $n+1$ is depicted by dashed symbols. (b) The 2D Brillouin zone.}
\label{fig:unitcell}
\end{figure}

The unit cell vectors are, see Fig.~\ref{fig:unitcell}
\[
{\mathbf a}_1=\frac{a_0}{2}(3\ , \; \sqrt{3})\ , \; {\mathbf a}_2=
\frac{a_0}{2}(3\ , \; -\sqrt{3})\ , \; {\mathbf a}_3 ={\mathbf d}+{\bm \delta}_1,
\]
where $a_0$ is the distance between two carbon atoms while $d$ is the interlayer distance in the $z$ direction. The nearest neighbor in-plane vectors are
\begin{equation}
{\bm \delta}_1 = a_0(1\ , 0)\ , \; {\bm \delta}_2=\frac{a_0}{2}(-1\ , \; \sqrt{3})\ , \; {\bm
\delta}_3=\frac{a_0}{2}(-1\ , \; -\sqrt{3})\ . \label{nn-vectors}
\end{equation}

The single-layer graphene has Dirac points in the Brillouin zone at 
\[
{\mathbf K} =\frac{2\pi }{3\sqrt{3}a_0}(\sqrt{3}\ , \; 1)\ , \;{\bf
K}^\prime =\frac{2\pi }{3\sqrt{3}a_0}(\sqrt{3}\ , \; -1)\ .
\]
In the vicinity of these points the graphene spectrum has a form of touching cones (for details, see review \cite{reviewCastroNeto09} and references therein). We show below that, for surface states in multilayered graphene with rhombohedral stacking as shown in Fig.~\ref{fig:rhggraphite}, the conical spectrum near the Dirac points is transformed into low-dispersion low-energy bands that determine the unique features of this system.

Therefore, we are interested here in low energies and thus in momenta close to one of the two non-equivalent Dirac corners ${\mathbf K}$ or ${\mathbf K}^\prime$ in the Brillouin zone. Let us expand the wave functions near
${\mathbf K}$ ,
\begin{eqnarray}
\psi_n^A({\mathbf R}_i)&=&\frac{1}{\sqrt{L}}\sum _{{\mathbf p}}  e^{i({\bf
K}+{\mathbf p}/\hbar)\cdot {\mathbf R}_i} \psi_n^A({\mathbf p})  \label{2-exp}\\
\psi_n^B ({\mathbf R}_i+{\bm \delta})&=&\frac{1}{\sqrt{L}}\sum
_{\tilde{\mathbf p}}  e^{i({\mathbf K}+\tilde{\mathbf p}/\hbar)\cdot ({\bf
R}_i+{\bm \delta})} \psi_n^B(\tilde{\mathbf p}) \ , \label{1-exp}
\end{eqnarray}
and assume that $|{\mathbf p}|, | \tilde{\mathbf p}| \ll \hbar a_0^{-1}$. Here $L$ is the number of unit cells in the plane (we omit the spin index). Small vectors ${\mathbf p}$ cannot couple the two Dirac points ${\mathbf K}$ and ${\mathbf K}^\prime$. Therefore, equations for them separate.

A standard Fourier series expansion of Hamiltonian \eqref{eq:tbbilayer}  near ${\mathbf K}$ yields 
\begin{eqnarray}
H_{{\mathbf K}}&=&\sum_{{\mathbf p}}\sum_{m,n=1}^N \hat \psi_m^\dagger ({\mathbf p}) \hat H_{mn}({\mathbf K},{\mathbf p}) \hat \psi_n  ({\mathbf p})\ ,\\  
\hat H_{mn}({\mathbf K},{\mathbf p})&=&\sum_{l=0}^4 \hat H_{mn}^{(l)}({\mathbf K},{\mathbf p}) \ , \label{H-particle}
\end{eqnarray}
where
\begin{eqnarray*}
\hat H_{mn}^{(0)}({\mathbf K},{\mathbf p})&=& v_F (\hat {\bm \sigma}\cdot {\mathbf p})\delta_{mn} \\
\hat H_{mn}^{(1)}({\mathbf K},{\mathbf p})&=& -\gamma_1 \left[ e^{-i\pi /6} \hat \sigma_+ \delta_{m,n+1}+  e^{i\pi /6} \hat \sigma_- \delta_{m,n-1}\right]  \\
\hat H_{mn}^{(3)}({\mathbf K},{\mathbf p})&=& \frac{\gamma_3}{\gamma_0}v_F \left[e^{-i\pi /3}   (\hat \sigma_+ p_+ ) \delta_{m,n-1} \right.\nonumber \\
&&+\left. e^{i\pi /3}  (\hat \sigma_- p_- ) \delta_{m,n+1}  \right]\\
\hat H_{mn}^{(4)}({\mathbf K},{\mathbf p})&=& \frac{\gamma_4}{\gamma_0} v_F \left[e^{i\pi /6}   p_-  \delta_{m,n-1} + e^{-i\pi /6}  p_+ \delta_{m,n+1} \right]
\end{eqnarray*}
and
$
v_F= 3a_0\gamma_0/2\hbar
$.
We define the pseudo-spinor
\[
\hat \psi_n =\left(\begin{array}{c} \psi_n^1 \\ \psi_n^2\end{array}\right)\ , \;
\hat \psi_n^\dagger =\left(  \psi_n^{1\dagger}\ , \; \psi_n^{2\dagger}\right)\ ,
\]
where $\psi^1_n = \psi_n^A $, $\psi^2_n=e^{i\pi/6} \psi_n^B$. Above we use
\[
2\hat \sigma_\pm =\hat \sigma_x \pm i\hat \sigma_y\ , \; p_\pm = p_x \pm i p_y =pe^{\pm i\phi}\ .
\]

We also need the Hamiltonian expansion near the opposite Dirac point $-{\mathbf K}$,
\begin{eqnarray}
\psi_n^A({\mathbf R}_i)&=&\frac{1}{\sqrt{L}}\sum _{{\mathbf p}}  e^{i(-{\bf
K}+{\mathbf p}/\hbar)\cdot {\mathbf R}_i} \bar \psi_n^A({\mathbf p})  \label{2-exp-}\\
\psi_n^B ({\mathbf R}_i+{\bm \delta})&=&\frac{1}{\sqrt{L}}\sum
_{\tilde{\mathbf p}}  e^{i(-{\mathbf K}+\tilde{\mathbf p}/\hbar)\cdot ({\bf
R}_i+{\bm \delta})} \bar \psi_n^B(\tilde{\mathbf p})\ . \label{1-exp-}
\end{eqnarray}
Using the same type of derivations we find 
\begin{equation}
H_{-{\mathbf K}}=\sum_{l,{\mathbf p}}\sum_{m,n=1}^N \hat {\bar \psi}_m^\dagger ({\mathbf p}) \hat H_{mn}^{(l)}(-{\mathbf K},{\mathbf p}) \hat {\bar \psi}_n  ({\mathbf p})\ , \label{H-hole}
\end{equation}
where $\hat H_{mn}^{(1)}(-{\mathbf K},{\mathbf p})=  \hat H_{mn}^{(1)*}({\mathbf K},{\mathbf p})$ and
$\hat H_{mn}^{(l)}(-{\mathbf K},{\mathbf p})=-  \hat H_{mn}^{(l)*}({\mathbf K},{\mathbf p})$ for $l\ne 1$. The Hamiltonian Eq. (\ref{H-hole}) at $-{\mathbf K}$ together with the Hamiltonian Eq. (\ref{H-particle}) at ${\mathbf K}$ are used below to construct the associated Bogoliubov-de Gennes Hamiltonian to describe the superconducting state.

In the next section we find numerically the energies and eigenstates of Hamiltonian Eq.~(\ref{H-particle}) using the relative magnitudes of the coupling constants listed above. The result of such numerics is displayed in Figs.~\ref{fig:3dspectrum} and \ref{fig:2dnormalspectrum}, along with the corresponding analytical approximations.

\subsection{Low-energy spectrum in the normal state}

The Schr{\"o}dinger equation takes the form
\begin{equation}
\sum _m \hat H_{nm}({\mathbf K},{\mathbf p}) \hat \psi_m ({\mathbf p}) =(\epsilon +\mu) \hat \psi_n({\mathbf p})\ . \label{Schr-eq1}
\end{equation}
The energy $E=\mu +\epsilon$ is measured from the chemical potential $\mu$. We use the Ansatz
\[
\hat \psi_n = \left(\begin{array}{c} A^1 \\ A^2e^{i\pi /6}\end{array}\right) e^{iqdn} \ ,
\]
where $q$ is the out-of-plane momentum, and obtain for $n\ne 1,N$
\begin{eqnarray}
\left[ v_F p\, e^{-i\tilde \phi} -\gamma_1 e^{-i qd}+ \frac{\gamma_3}{\gamma_0}v_Fp \, e^{i(\tilde \phi +qd)}\right]   A^2 \nonumber \\
+\left[ 2\frac{\gamma_4}{\gamma_0} v_Fp \cos(\tilde \phi -qd) -(\epsilon +\mu)\right] A^1 =0\nonumber \\
\left[ v_F p\, e^{i\tilde \phi} -\gamma_1 e^{i qd}+ \frac{\gamma_3}{\gamma_0}v_Fp \, e^{-i(\tilde \phi +qd)}\right] A^1 \nonumber \\
+\left[ 2\frac{\gamma_4}{\gamma_0} v_Fp \cos(\tilde \phi -qd) -(\epsilon +\mu)\right]   A^2 =0 \ ,
\label{bulksp}
\end{eqnarray}
with $\phi = \tilde \phi  +\pi/6$. The compatibility condition yields the energy spectrum for excitations in the bulk,
\begin{eqnarray*}
\left[ \epsilon +\mu - 2\frac{\gamma_4}{\gamma_0} v_Fp \cos(\tilde \phi -qd)\right]^2
= v^2p^2 +\gamma_1^2  -2(vp)\gamma_1 \cos(\tilde \phi -qd)\nonumber \\
+\left(\frac{\gamma_3}{\gamma_0}\right)^2(vp)^2 
 +2\frac{\gamma_3}{\gamma_0}(vp)^2 \cos(2\tilde \phi +qd)-2\gamma_1\frac{\gamma_3}{\gamma_0}(vp)\cos(\tilde \phi+2qd)\ . 
\end{eqnarray*}

\begin{figure}[h]
\centering
\includegraphics[width=8cm]{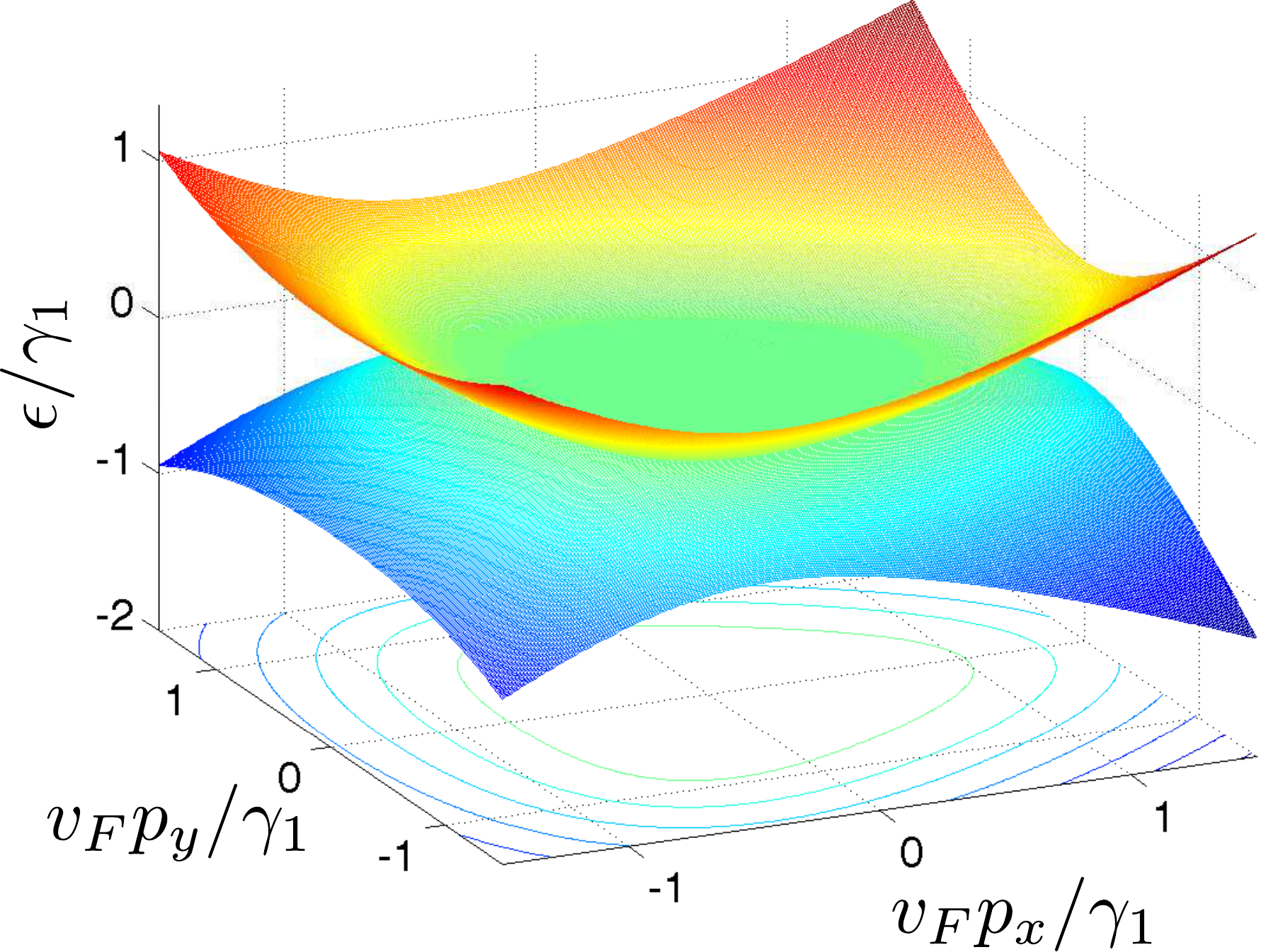}
\caption{Spectrum of rhombohedral graphite surface state as a function of momenta $p_x$ and $p_y$. On this scale, the effect of $\gamma_4$ does not show up. Due to the non-zero $\gamma_3$, the shape of the low-energy region is distorted from a circle into a more triangular shape.}
\label{fig:3dspectrum}
\end{figure}

For zero doping $\mu =0$, the Fermi surface is determined by
$\epsilon(p,q,\phi)=0$. If $\gamma_3 = \gamma_4 =0$, the Fermi surface shrinks to the spiral
$
v_Fp= \gamma_1\ , \; \tilde\phi =qd
$.
If only $\gamma_4 =0$, the Fermi surface for zero doping is determined by
\begin{eqnarray}
\frac{v_F p}{\gamma_1}
=\frac{e^{\pm i(\tilde \phi -qd)} +(\gamma_3/\gamma_0)e^{\mp i(\tilde \phi +2qd)}} {1+(\gamma_3/\gamma_0)^2 +2(\gamma_3/\gamma_0)\cos(2\tilde \phi+qd)}\ . \label{Fermi-surf}
\end{eqnarray}
\begin{figure}[h]
\centering
\includegraphics[width=8cm]{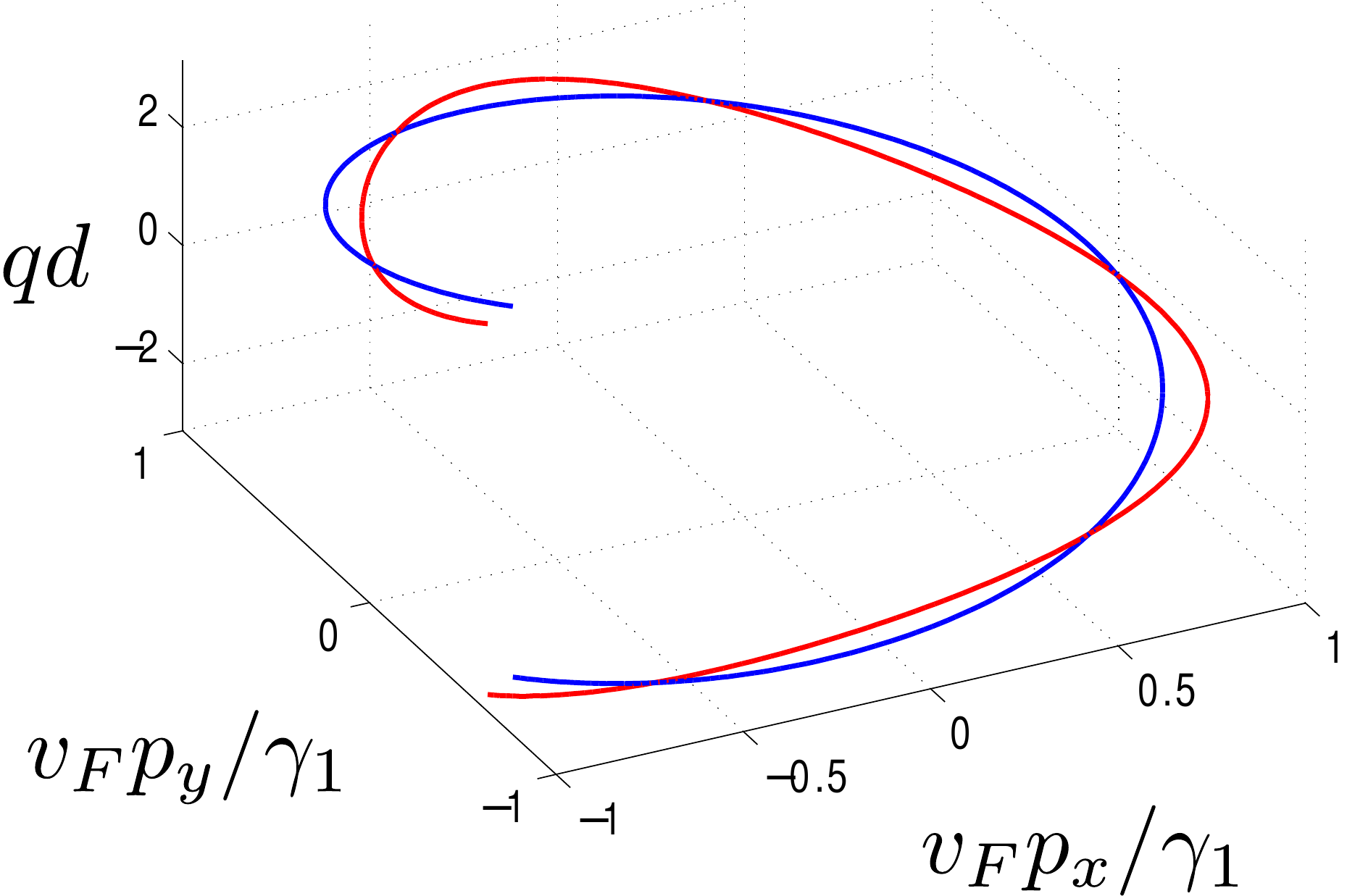}
\caption{Fermi line of rhombohedral graphite for $\gamma_4=0, \gamma_3\neq 0$ (red) compared to the simple spiral obtained for $\gamma_4=\gamma_3=0$ (blue).}
\label{fig:spirals}
\end{figure}
Real momenta $p$ and $q$ are realized along the line $q(\phi)$ satisfying
\[
\sin(\tilde \phi-qd)-(\gamma_3/\gamma_0)\sin(\tilde \phi+2qd)=0
\]
or
\[
\tan \tilde \phi = \frac{\sin (qd) +(\gamma_3/\gamma_0)\sin (2qd)}{\cos (qd) -(\gamma_3/\gamma_0)\cos (2qd)} \ .
\]
Along this line
\[
\frac{v_Fp}{\gamma_1}=\frac{\cos(dq -\tilde \phi)+\gamma_3 \cos(2qd+\tilde \phi)}{1+\gamma_3^2+2\gamma_3 \cos(qd+2\tilde \phi)}\ .
\]
It is a corrugated spiral in the 3D space $(p,q,\phi)$. 
Since $\tilde \phi$ at the Fermi line as a function of $qd$ is periodic with period $2\pi$, the corrugation has a three-fold symmetry in the plane of $\phi$, as is seen in Fig.~\ref{fig:spirals} and in Fig.~\ref{fig:3dspectrum} below for the corresponding flat band. It is clear that the $\gamma_3$ interaction does not destroy the Fermi line. Indeed, this is because the interaction comes with matrices $\hat \sigma_x$ and $\hat \sigma_y$  and thus the full Hamiltonian obeys the same anti-commutation rule 
\[
\left[ \hat \sigma_z ,(\hat H^{(0)} +\hat H^{(1)}+ \hat H^{(3)})\right]_+ =0
\]
as the initial Hamiltonian $ \hat H^{(0)} +\hat H^{(1)}$.  According to \cite{HeikkilaVolovik10-1}, this preserves the same topological invariant and hence the topology of the Fermi surface is unchanged. 

We now turn to the surface states with low energies $\epsilon =0$. A surface state corresponds to a complex out-of-plane momentum $q=q^\prime +i q^{\prime\prime}$ that ensures its decay into the bulk. 
Since the interaction proportional to $\gamma_3$ does not change the conclusion about the presence of a flat band, we restrict our analytical consideration to the case when only $\gamma_4$ is nonzero while $\gamma_3=0$ for simplicity. Note that our  numerical analysis is carried out using the exact diagonalization of the full Hamiltonian (with non-zero $\gamma_3$). For $\gamma_3=0$ Eq. (\ref{Schr-eq1}) takes the form
\begin{eqnarray}
v_F  (\hat {\bm \sigma}\cdot {\mathbf p}) \hat \psi_n ({\mathbf p}) -\gamma_1 \left[ e^{i\pi /6} \hat \sigma_- \hat \psi_{n+1} + e^{-i\pi /6} \hat \sigma_+ \hat \psi_{n-1}\right] \nonumber \\
+\frac{\gamma_4}{\gamma_0} \left[ e^{i\pi /6} v_Fp_- \hat \psi _{n+1} +  e^{-i\pi /6} v_Fp_+ \hat \psi _{n-1}\right] =(\epsilon +\mu)\hat \psi_n \ .\label{Schr-eq2}
\end{eqnarray}
Equation (\ref{Schr-eq2}) suggests a solution in the form 
\begin{eqnarray}
\hat \psi_n &=&  e^{i(\phi -\pi/6)(n-1-\frac{N}{2})}\nonumber \\
&&\times \left[\left(\frac{vp}{\gamma_1}\right)^{n-1}\left(\begin{array}{c} 1 \\ \zeta e^{i\phi} \end{array}\right)A_+ + \left(\frac{vp}{\gamma_1}\right)^{N-n}\left(\begin{array}{c} \zeta \\ e^{i\phi}\end{array}\right)A_-\right]\ , \qquad \label{psi}
\end{eqnarray}
where
\[
\zeta = \left(\frac{vp}{\gamma_1}\right)\frac{\gamma_1 (\epsilon +\mu)-(\gamma_4/\gamma_0)(v^2p^2+\gamma_1^2)}{v^2p^2-\gamma_1^2}\ .
\]
This form of solution implies the out-of-plane momentum $q^\prime d =\tilde \phi$ while $e^{\pm q^{\prime\prime}d} =(v_Fp/\gamma_1)$.
\begin{figure}[h]
\includegraphics[width=8cm]{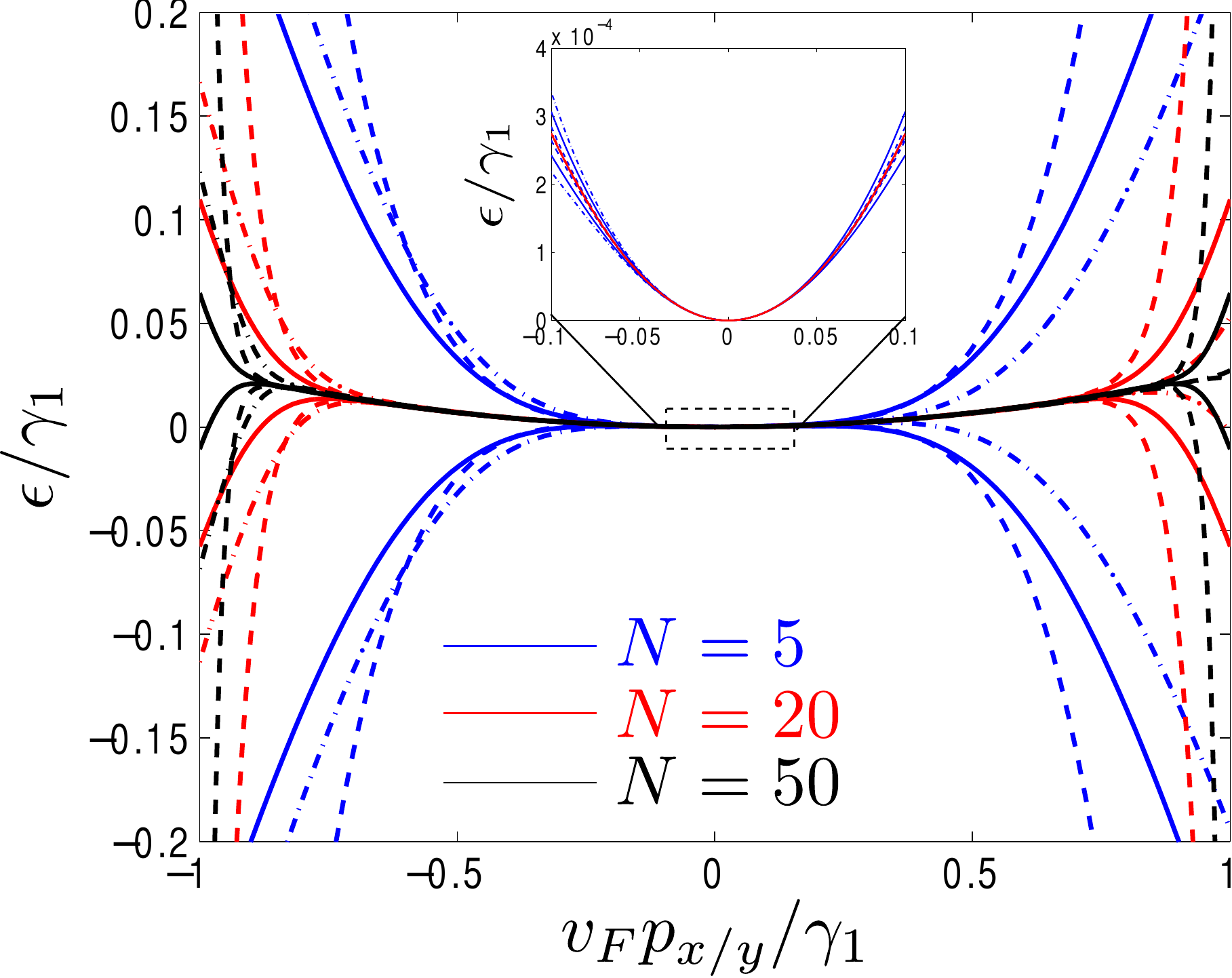}
\caption{Cuts of the 3d spectrum shown in Fig.~\ref{fig:3dspectrum} along the $p_x$ (solid lines) and $p_y$ (dash-dotted lines) directions and limited to the low-energy region $v_F p<\gamma_1$. Three cases are shown, $N=5$ (blue), $N=20$ (red), and $N=50$ (black) graphene layers. The inset shows a zoom-up of the low-energy region. The dashed lines show the corresponding approximations from Eq.~\eqref{eq:normaldispersion}, without including the correction $1-v^2p^2/\gamma_1^2$ as this becomes relevant only in the superconducting state. The deviations between the dashed and other lines show up mostly when the term $\xi_p$ becomes dominant in Eq.~\eqref{eq:normaldispersion}, and are partially due to the term $\gamma_3$ neglected in that approximation.}
\label{fig:2dnormalspectrum}
\end{figure}

At the outermost layers $n=1,N$, the terms which would contain $\hat \psi_0$ and $\hat \psi_{N+1}$ in Eq.~(\ref{Schr-eq2}) disappear.
Those components for which the terms with $\gamma_1$ are absent have the form
\begin{eqnarray}
v_F  pe^{-i\phi} \psi_1^2 
+\frac{\gamma_4}{\gamma_0} \left[ v_Fp e^{-i\phi+i\pi /6} \psi _{2}^1 \right] =(\epsilon +\mu) \psi_1^1 \\
v_F  pe^{i\phi}\psi_N^1 
+\frac{\gamma_4}{\gamma_0} \left[ v_Fp e^{i\phi -i\pi /6}\psi _{N-1}^2\right] =(\epsilon +\mu) \psi_N^2 \ . \label{schr-surf1}
\end{eqnarray}
They couple the constants $A^+$ and $A^-$ and determine the energy of the surface states. We define
\[
\xi_p=\gamma_1 \left( v_Fp/\gamma_1 \right)^N\ .
\]
Using Eq. (\ref{psi}) we find for $\xi_p , \epsilon \ll \gamma_1$
\begin{eqnarray*}
\xi_p A_\pm  =\left[ \frac{\gamma_1^2}{\gamma_1^2-v^2p^2} (\epsilon +\mu) - 2\gamma_1\frac{\gamma_4}{\gamma_0}\frac{v^2p^2}{\gamma_1^2 -v^2p^2}\right] A_\mp\ .
\end{eqnarray*}
The normal-state dispersion is $\epsilon =\epsilon_p$,
\begin{equation}
\epsilon_p = \mu_p  \pm \xi_p \left(1-v^2p^2/\gamma_1^2\right)\ , \;
\mu_p= \frac{p^2}{2m^*} -\mu\ ,
\label{eq:normaldispersion}
\end{equation}
where
\begin{equation}
m^*=\frac{\partial ^2 \epsilon_p}{\partial p^2}=\frac{\gamma_1\gamma_0}{4 \gamma_4 v_F^2}\ . \label{mass}
\end{equation}
The spectrum $\epsilon_p$ has a quadratic dispersion with the effective mass $m^*$ on a background of a much weaker high-order dispersion $\xi_p$. The latter transforms into a flat band $\xi_p=0$ with a radius $v_Fp<\gamma_1$ for an infinite number of layers, $N\to \infty$. This form of the dispersion is compatible with the findings made with the numerical calculations exploiting density functional theory:\cite{KopninIjasHarjuHeikkila12} also there quadratic dispersion with an effective mass of the same order as that estimated here is found.

The effective mass is much larger than the characteristic band mass $m_3$ in 3D graphite. Indeed, we
have
\[
\frac{m^*}{m_3}\sim \frac{\gamma_1 \hbar^2}{ \gamma_4 \gamma_0 m_3 a_0^2} \sim \frac{\gamma_1 }{ \gamma_4}\ ,
\]
where we estimate $\hbar^2/(m_3 a_0^2) \sim \gamma_0$ as the conduction band width in graphite. We see that $m^*/m_3 \gg 1$. 
The group velocity is $v_g=\partial \epsilon_p/\partial p = p/m^*$.

This dispersion is compared with the exact dispersion obtained from the numerical diagonalization of $H({\mathbf K}, {\mathbf p})$ in Figs.~\ref{fig:3dspectrum} and \ref{fig:2dnormalspectrum} using $\gamma_3/\gamma_0=0.098$, and $\gamma_4/\gamma_0=0.014$ according to \cite{reviewCastroNeto09,Dresselhaus02}.

\section{Bogoliubov-de Gennes equations for the superconducting state} \label{sec:bdgequations}

To construct the BdG equations we need the Hamiltonian for holes.\footnote{Note the difference between BdG holes that are time-reversed electron states and the valence band excitations (absence of electrons below the Fermi level), which are sometimes also referred as holes. By construction, the BdG equations are electron-hole symmetric.} The hole Hamiltonian near a Dirac point ${\mathbf K}$ follows from the particle Hamiltonian in a vicinity of the opposite Dirac point $-{\mathbf K}$.
The wave function $\psi_{{\mathbf K}}^h $ of a hole excitation near the Dirac point ${\mathbf K}$ is  $\psi_{{\mathbf K}}^h = \bar \psi_{-{\mathbf K}}^*$. Therefore, the hole Hamiltonian is $H^h({\mathbf K},{\mathbf p})=H^*(-{\mathbf K},-{\mathbf p})$. Since the term $H_{mn}^{(1)}({\mathbf K},{\mathbf p})$ is independent of ${\mathbf p}$ while the other terms are linear in ${\mathbf p}$ we have using Eq. (\ref{H-hole})
\begin{eqnarray*}
H^h_{{\mathbf K}} &=&\sum_{l,{\mathbf p}}\sum_{m,n=1}^N \hat {\psi}_m^{(h) \dagger } ({\mathbf p}) \hat H_{mn}^{(l)*}(-{\mathbf K},- {\mathbf p}) \hat {\psi}_n^{(h)}  ({\mathbf p}) \\
&=& \sum_{l,{\mathbf p}}\sum_{m,n=1}^N \hat {\psi}_m^{(h) \dagger } ({\mathbf p}) \hat H_{mn}^{(l)}({\mathbf K},{\mathbf p}) \hat {\psi}_n^{(h)}  ({\mathbf p})\\
&=& \sum_{{\mathbf p}}\sum_{m,n=1}^N \hat {\psi}_m^{(h) \dagger } ({\mathbf p}) \hat H_{mn}({\mathbf K},{\mathbf p}) \hat {\psi}_n^{(h)}  ({\mathbf p})\ .
\end{eqnarray*}
In other words, the hole Hamiltonian coincides with Eq. (\ref{H-particle}) where the electronic wave functions are replaced with the corresponding hole functions. 
As distinct from the quasiparticle energy measured from the chemical potential upwards, $E=\mu +\epsilon$, the energy of holes is measured from the chemical potential downwards, $E=\mu - \epsilon$.
In what follows, the electron wave function is denoted by $\hat u_n = \hat \psi_n$  while the hole wave function is denoted by $\hat v_n =\hat \psi^h_n$.

The BdG equations for the superconducting state are constructed using the Schr{\"o}dinger equations of the type of Eq. (\ref{Schr-eq2}) for particles and holes with the particle-hole coupling through the order-parameter field $\Delta$,
\begin{eqnarray}
\sum_m \check \tau_3 \otimes \left[ \hat H_{nm}({\mathbf K},{\mathbf p}) -\mu  \delta_{nm}\right]\check \Psi_m + \check \Delta_n \check \Psi_n =\epsilon \check \Psi_n\ . \label{BdGHamilt}
\end{eqnarray} 
Here we introduce objects in the Nambu space
\[
\check \tau_3=\left(\begin{array}{lr} 1 & 0 \\ 0&
-1\end{array}\right)\ , \; \check \Delta_n
=\left(\begin{array}{lr} 0 & \Delta_n \\ \Delta^*_n &
0\end{array}\right)\ , \; \check \Psi_n =
\left(\begin{array}{c} \hat u_n 
\\ \hat v_n   \end{array}\right)\ .
\]
The Nambu vector $\check \Psi_n$ 
has the pseudo-spinor components  
\[
\hat u_n =\left(\begin{array}{c} u_n^1\\ u_n^2\end{array}\right)\ , \;
\hat v_n =\left(\begin{array}{c} v_n^1\\ v_n^2\end{array}\right) \ .
\]

For $\gamma_3 =0$ the BdG equations in components take the form (for $n \ne 1,N$)
\begin{eqnarray}
\left[ v_F  (\hat {\bm \sigma}\cdot {\mathbf p}) \hat u_n ({\mathbf p}) -\gamma_1 \left( e^{i\pi /6} \hat \sigma_- \hat u_{n+1} + e^{-i\pi /6} \hat \sigma_+ \hat u_{n-1}\right) \right. \nonumber \\
+\left. \frac{\gamma_4}{\gamma_0} \left( e^{i\pi /6} v_Fp_- \hat u_{n+1} +  e^{-i\pi /6} v_Fp_+ \hat u _{n-1}\right) -\mu \hat u_n \right] \nonumber \\
+\Delta_n \hat v_n =\epsilon \hat u_n\ , \label{BdG-eq-u} \\
-\left[v_F  (\hat {\bm \sigma}\cdot {\mathbf p}) \hat v_n ({\mathbf p}) -\gamma_1 \left( e^{i\pi /6} \hat \sigma_- \hat v_{n+1} + e^{-i\pi /6} \hat \sigma_+ \hat v_{n-1}\right) \right. \nonumber \\
+\left. \frac{\gamma_4}{\gamma_0} \left( e^{i\pi /6} v_Fp_- \hat v_{n+1} +  e^{-i\pi /6} v_Fp_+ \hat v _{n-1}\right)-\mu \hat v_n\right]\nonumber \\
+\Delta_n^* \hat u_n = \epsilon \hat v_n \ . \label{BdG-eq-v}
\end{eqnarray}

At the outermost layers $n=1,N$, the terms with $\hat u_0, \ \hat v_0$ and $\hat u_{N+1}, \ \hat v_{N+1}$  in Eqs. (\ref{BdG-eq-u}), (\ref{BdG-eq-v}) disappear.
For those components which do not contain the terms with $\gamma_1$ we have 
\begin{eqnarray}
 v_F  pe^{-i\phi} \check \tau_3 \check \alpha_1^- + \frac{\gamma_4}{\gamma_0}  v_Fp e^{-i\phi+i\pi /6}\check \alpha_2^+ -\mu \check \alpha_1^+ + \check \Delta_1 \check \alpha_1^+ &=& \epsilon \check \alpha_1^+ \ , \qquad \label{BdG-surf-a+} \\
 v_F  pe^{i\phi} \check \tau_3 \check \alpha_N^+ + \frac{\gamma_4}{\gamma_0}  v_Fp e^{i\phi-i\pi /6}\check \alpha_{N-1}^- -\mu \check \alpha_N^- + \check \Delta_N \check \alpha_N^- &=& \epsilon \check \alpha_N^- \ . \qquad \label{BdG-surf-a-} 
\end{eqnarray}
Here we decompose the wave function
\begin{equation}
\check \Psi _n =\left[ \left(\begin{array}{c} \alpha_n^+
\\ \beta_n^+ \end{array}\right) \otimes \hat \psi^+  +
\left(\begin{array}{c} \alpha_n^-
\\ \beta_n^- \end{array}\right) \otimes \hat \psi^- \right]\label{wavefunc-u}
\end{equation}
into the spinor functions localized at each sublattice
\[
\hat \psi^+   =\left(\begin{array}{c} 1\\0\end{array}\right)
 \ , \; \hat \psi^-   =\left(\begin{array}{c}
0\\1\end{array}\right)  
\]
and introduce the vector in the Nambu space
\[
\check \alpha_n^\pm =
\left(\begin{array}{c} \alpha_n^\pm
\\ \beta_n^\pm  \end{array}\right)\ .
\]

For analytical consideration we assume that $\Delta_n =0$ for $n\ne 1,N$. This assumption is justified by the results of numerical solution of the self-consistency equation with the wave functions found from the full BdG equations, discussed below, see especially Fig.~\ref{fig:Deltaprofile}. 
In this case, Eqs.~(\ref{BdG-eq-u})--(\ref{BdG-eq-v}) for $n\ne 1,N$ do not contain $\Delta$, so that one can use the
normal-state coefficients as in Eq. (\ref{psi}), 
\begin{eqnarray*}
\check \alpha^+_n &=&\frac{C}{\sqrt{2}}e^{i(n-1-\frac{N}{2})(\phi-\frac{\pi}{6})}\left[ \left(
\frac{v_Fp}{\gamma_1}\right)^{n-1}\check A^+ + \left( \frac{v_Fp}{\gamma_1}\right)^{N-n}\check \zeta  \check A^-\right]\ , 
\\
\check \alpha^-_n &=&\frac{C}{\sqrt{2}}e^{i(n-1-\frac{N}{2})(\phi-\frac{\pi}{6})}\left[\left(
\frac{v_Fp}{\gamma_1}\right)^{N-n} \check A^- + \left( \frac{v_Fp}{\gamma_1}\right)^{n-1}\check \zeta  \check A^+\right]e^{i\phi}\ . 
\end{eqnarray*}
Here $C$ is a normalization constant; the vectors $ \check A^\pm =
\left(A^\pm , \, B^\pm \right)^T$ do not depend on $n$. We also define the matrix
\[
\check \zeta  = \left(\frac{vp}{\gamma_1}\right)\frac{\gamma_1 (\check \tau_3 \epsilon +\mu)-(\gamma_4/\gamma_0)(v^2p^2+\gamma_1^2)}{v^2p^2-\gamma_1^2}\ .
\] 

Equations (\ref{BdG-surf-a+})--(\ref{BdG-surf-a-}) yield
\begin{eqnarray}
\check \tau_3 \xi_{ p} \check A^-= (\tilde \epsilon - \check \tau_3 \tilde
\mu_p )\check A^+ - \check
\Delta_1 \check A^+ \ , \label{BdGeq-mu1}\\
\check \tau_3 \xi_{ p} \check A^+= (\tilde \epsilon - \check \tau_3 \tilde
\mu_p )\check A^- - \check \Delta_N \check A^- \ ,
\label{BdGeq-mu4}
\end{eqnarray}
where
\[
\tilde \epsilon =\epsilon \left(1-v^2p^2/\gamma_1^2\right)^{-1}\ , \;
\tilde \mu_p =\mu_p \left(1-v^2p^2/\gamma_1^2\right)^{-1}\ .
\]

\begin{figure}[h]
\centering
\includegraphics[width=8cm]{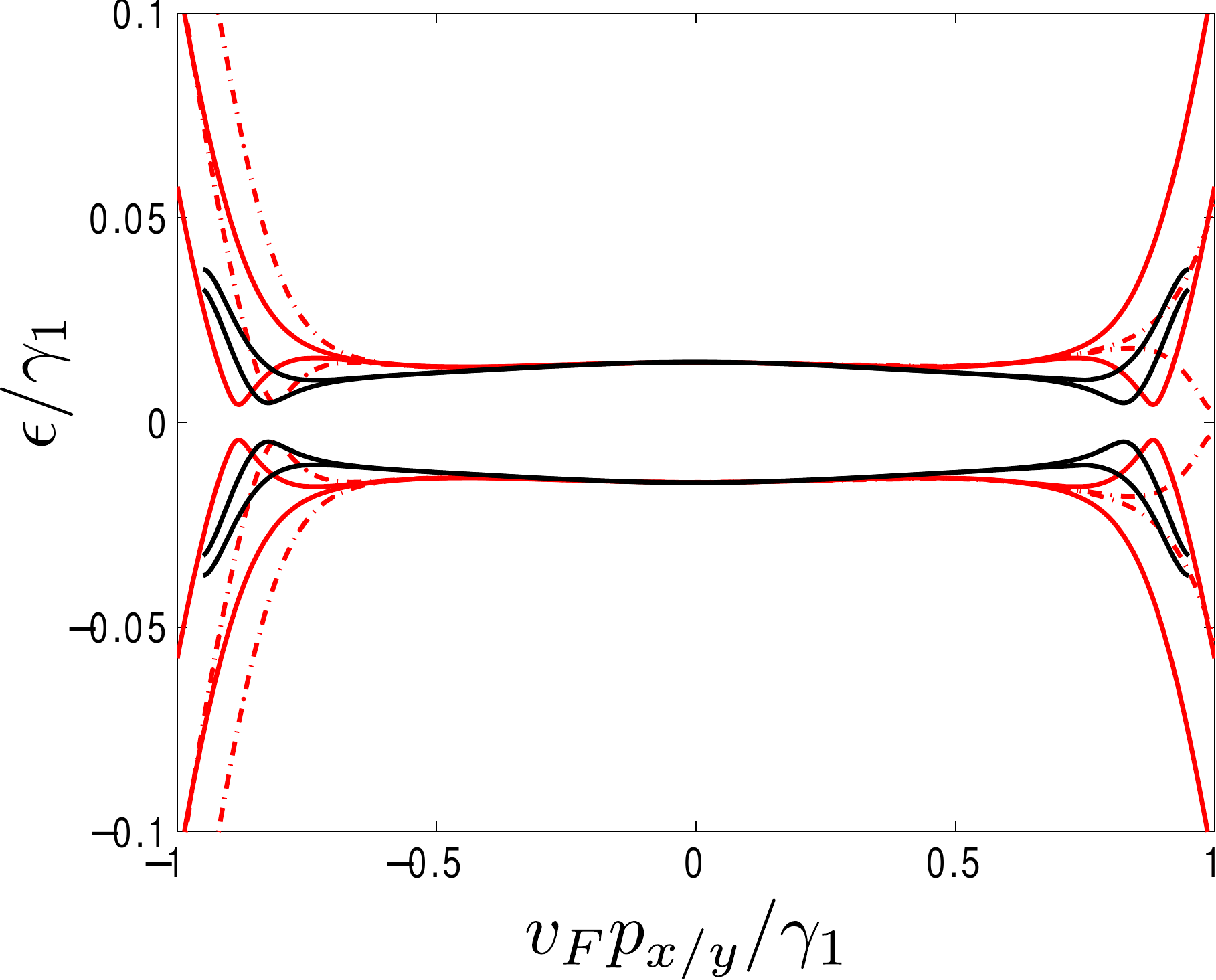}
\caption{Spectrum in the superconducting state, assuming a coupling constant $g=\gamma_1/2$ yielding a gap $\Delta_1=\Delta_N=0.015 \gamma_1$ for $N=20$ layers. The Bogoliubov-de Gennes equation has four branches of solutions, two for the electrons and two for the holes. The direction $p_y=0$ is shown with the solid lines and $p_x=0$ with the dash-dotted lines. The black lines are the corresponding approximations from Eq.~\eqref{eq:bdgspectrum} and the deviations with the exact numerics are mostly due to the $\gamma_3$ term neglected in Eq.~\eqref{eq:bdgspectrum}.}
\label{fig:bdgspectrum}
\end{figure}

Equations (\ref{BdGeq-mu1}), (\ref{BdGeq-mu4}) provide the
surface-state spectrum \cite{KopninHeikkilaVolovik2011}
\begin{eqnarray}
\left[ \tilde \epsilon ^2-\tilde \mu_p ^2-|\Delta_N|^2\right]
\left[\tilde \epsilon ^2-\tilde\mu_p ^2-|\Delta_1|^2\right]+\xi_{ p}^4 \nonumber \\
-\xi_{ p}^2\left[2\tilde \epsilon ^2 +2\tilde \mu_p^2  
-\Delta_1^*\Delta_N - \Delta_1\Delta_N^*\right]=0 \ .
\label{cond1}
\end{eqnarray}
If $\Delta_1 =\Delta_N$ we have
\begin{equation}
\tilde \epsilon^2 = \left(\tilde \mu_p \pm  \xi_p\right)^2+|\Delta|^2.
\label{eq:bdgspectrum}
\end{equation}
This is compared to the exact diagonalization of the Bogoliubov-de Gennes equations in Fig.~\ref{fig:bdgspectrum}.

Equations (\ref{BdGeq-mu1}), (\ref{BdGeq-mu4}) determine four independent states. If $\gamma_4=\mu=0$ they are:
(i) $\tilde \epsilon_1=\tilde E$ and $A_1^\pm = u
$, $B_1^\pm = v $, (ii) $\tilde \epsilon_2=- \tilde E$ and $A_2^\pm = v
$, $B_2^\pm =- u $, (iii) $\tilde \epsilon_3=\tilde E$ and $A_3^\pm
=\pm   v $, $B_3^\pm =\pm  u$, (iv) $\tilde \epsilon_4=-\tilde E$ and
$A_4^\pm =\pm u $, $B_1^\pm =\mp v $. Here $\tilde E=
\sqrt{\xi_p^2 +\Delta^2}$ and
\begin{equation}
u=\frac{1}{\sqrt{2}}\left[ 1+ \xi_{ p}/\tilde E
\right]^\frac{1}{2} , \; v=\frac{1}{\sqrt{2}}\left[ 1- \xi_{
p}/\tilde E \right]^\frac{1}{2}\ .\; \label{A}
\end{equation}
The overall normalization requires $d \sum_{n=1}^N[|\alpha^+_n|^2
+|\beta^+_n|^2+|\alpha^-_n|^2 +|\beta^-_n|^2]=1$. For $\xi_{ p}\ll
\gamma_1$ this gives
\begin{equation}
|C |^2=d^{-1}[1-(v_Fp/\gamma_1)^2]\ . \label{norm}
\end{equation}

If the number of layers $N$ is large, $\xi_p \to 0$ for $v_Fp<\gamma$, the two surface states decouple
\begin{equation}
\tilde \epsilon _1^2 =\tilde \mu_p^2 +|\Delta|^2_1 \ , \; \tilde \epsilon _N^2 =\tilde \mu_p^2 +|\Delta|^2_N \ .\label{spectr-flat1}
\end{equation}
In this case, Eq. (\ref{BdGeq-mu1}) yields
\begin{eqnarray}
(\tilde \epsilon - \check \tau_3 \tilde
\mu_p )\check A^+ - \check \Delta_1 \check A^+ =0
\end{eqnarray}
whence $A^+ = U $, $B^+ = V $ or $A^+ = V $, $B^+ = U $ where
\begin{equation}
U =\frac{1}{\sqrt{2}}\left[ 1 + \tilde \mu_p/\tilde \epsilon\right]^{\frac{1}{2}}\ ,\;  V=\frac{1}{\sqrt{2}}\left[ 1 -\tilde \mu_p/\tilde \epsilon \right]^{\frac{1}{2}}\ . \label{UV}
\end{equation}
In the following section we use these eigenfunctions to calculate the self-consistent gap function $\Delta$.

\section{Surface superconductivity}
\label{sec:superconductivity}

The surface states discussed above form a basis for the superconducting order parameter localized near outer surfaces.
Within the mean-field approximation, the order parameter at layer $n$ is determined by the self-consistency equation,
\begin{eqnarray}
\Delta_n&=& \int \frac{d^2 p}{(2\pi \hbar)^2}\sum_{q} 
W \, {\rm Tr}\, [\hat u_n({\mathbf p},k) \hat v^*_n({\mathbf p},k)]\nonumber \\
&&\times [1-2f(E_{{\mathbf p}, k})]\ , \label{OP1}
\end{eqnarray}
where $W$ is the 3D coupling potential, $f(E)$ is the Fermi distribution function. The sum includes all states $q$ with given 2D momentum ${\mathbf p}$. As was shown in \cite{KopninHeikkilaVolovik2011}, the terms corresponding to the surface states dominate in the sum due to a much larger density of states compared to the bulk states. For a large number of layers $N\to \infty$ when $\xi_p=0$, the self-consistency equation for the order parameter on the surface takes the form
\begin{eqnarray}
\Delta &=&2W \int_{p<p_{FB}}\frac{d^2 p}{(2\pi \hbar)^2} |C|^2 UV \tanh\frac{\epsilon}{2T}\ , \label{eq-self1}
\end{eqnarray}
where $p_{\rm FB}=\gamma_1/v_F$. Using $U$ and $V$ from Eq. (\ref{UV}),
\begin{eqnarray}
\Delta&=&\frac{W}{d} \int_{p<p_{FB}}\frac{d^2 p}{(2\pi \hbar)^2} \frac{\left(1- v^2p^2/\gamma_1^2 \right)^2 \Delta}{\sqrt{\mu_p^2+|\Delta|^2(1-v^2p^2/\gamma_1^2)^2}} 
\nonumber\\
&&\times \tanh\frac{\sqrt{\mu_p^2+|\Delta|^2(1-v^2p^2/\gamma_1^2)^2}}{2T}\ .\label{Delta-gen} 
\end{eqnarray}
In the following sections we consider several examples that can be derived from Eqs.~(\ref{OP1})--(\ref{Delta-gen}).

\subsection{Flat band}

For zero doping $\mu =0$, in the absence of a band curvature $\gamma_4=0$ and $\xi_p =0$, Eq. (\ref{Delta-gen}) yields the flat-band result (for $T=0$) \cite{KopninHeikkilaVolovik2011}
\begin{eqnarray}
\Delta 
=\frac{W}{d} \int_{p<p_{FB}}\frac{d^2 p}{(2\pi \hbar)^2} \left(1-\frac{v^2p^2}{\gamma_1^2}\right) =
\frac{g}{8\pi}\ , \label{Delta-FB}
\end{eqnarray}
where the coupling energy is
\[
g= \frac{(W/d)\gamma_1^2}{v^2\hbar^2}=\frac{(W/d)p_{FB}^2}{ \hbar^2}\ .
\]
The coupling energy can be expressed in terms of the usual BCS coupling constant $\lambda =\nu_3 W$ where $\nu_3 =m_3 p_{3F}/2\pi^3 \hbar^3$ is the 3D density of states. Assuming the conduction band width in 3D graphite of the order of $\gamma_0$ we have
\begin{equation}
g \sim \lambda  \frac{\gamma_1^2}{\gamma_0}\frac{\hbar}{p_{3F} a_0} \label{g-estim}\ ,
\end{equation}
which can be estimated as $g/\gamma_1 \sim \lambda (\gamma_1 /\gamma_0)$ if $ \hbar /a_0p_{3F} \sim 1$.

The critical temperature
is determined by Eq.~(\ref{Delta-gen}) with $\Delta \to 0$, which
gives $\Delta_0=3k_B T_c$. Due to its linear dependence on the
interaction strength, {\it the critical temperature is
proportional to the area of the flat band and can be essentially
higher than that in the bulk}.

Doping in the flat band regime destroys the surface superconductivity \cite{KopninHeikkilaVolovik2011}. This can be seen
from Eq.~(\ref{eq-self1}) with $UV= \Delta/2 \tilde \epsilon_1$ and
where $\tilde \epsilon_1$ is taken from
Eq.~(\ref{spectr-flat1}). Both $\Delta_0$ and $T_c$ vanish at the critical doping level  $|\mu| =2k_B T_c$.

For a flat band $\xi_{ p} =0$ with $p_c=p_{\rm FB}$ the only
characteristic values in the superconducting surface state are the
energy $\Delta$ and the momentum $p_{\rm FB}$. Therefore, the
coherence length should be of the order of the only available
length scale, $ \xi_0 \sim \hbar/p_{\rm FB} $. It is much larger
than the interatomic distance, $\xi_0 \gg a_0$, since $p_{\rm FB}
\ll p_0\sim \hbar /a_0$.

\subsection{``Flat-band'' surface superconductivity in a finite array.}\label{subsec:flatband}

Let us discuss the ``flat band'' regime $\mu=0$ and $\gamma_4=0$ for a system with a finite number of layers. Since the normal-state DOS defined as
\begin{equation}
\nu(\xi_p) = \frac{p}{2\pi\hbar^2}\,
\frac{dp}{d\xi_p}=\frac{\gamma_1(\xi_p /\gamma_1)^{\frac{2-N}{N}}}{2\pi \hbar^2
N v_F^2}\  \label{DOS}
\end{equation}
has a low-energy singularity for $N>2$, the surface
superconductivity is favorable already for a system with a finite
number of layers $N\geq 3$. A simple expression for the
zero-temperature gap can be obtained if $N\geq 5$. For a finite
$N$, the value $\xi_p$ can reach values larger than $\Delta$. We
insert Eqs. (\ref{A}) at zero doping for $u$ and $v$ together with Eq. (\ref{norm}) and Eq. (\ref{eq:bdgspectrum}) into
Eq.~(\ref{eq-self1}). Since the upper limit of integration is $p_{\rm FB}$,  the corresponding upper limit for $\xi_p$ is $\xi_c=\gamma_1 \gg \Delta$. Transforming to
the energy integral with the normal-state DOS Eq.~(\ref{DOS}) we find for $T=0$
\begin{eqnarray}
1&=&  \frac{W}{d} \int_0^{\xi_c} \nu(\xi_p)\,
\frac{1-(\xi_p/\gamma_1)^\frac{2}{N}}{\sqrt{\xi_p^2+|\Delta|^2}}\, d\xi_p
\nonumber \\
&=&\frac{W\gamma_1}{2\pi \hbar^2
d N v_F^2} \int_0^{\xi_c} \frac{(\xi_p/\gamma_1)^\frac{2-N}{N}
[1-(\xi_p/\gamma_1)^\frac{2}{N} ]}{\sqrt{\xi_p^2+|\Delta|^2}}\,
d\xi_p  \ . \label{DeltavsN}
\end{eqnarray}

We see that, for $N>4$, the integral converges at $\xi_p \sim \Delta$
or $p\sim p_\Delta =p_{\rm FB}(\Delta/\gamma_1)^\frac{1}{N}$. The
zero-temperature gap is
\begin{equation}
\Delta_0 =\gamma_1\left(\frac{ g }{4\pi
\gamma_1}\left[\alpha(N)-\frac{1}{2}\left(\Delta_0/\gamma_1\right)^\frac{2}{N}
\alpha(N/2)\right] \right)^\frac{N}{N-2}
\end{equation}
where
\[
\alpha(N) =\int_0^{\infty} \frac{x^\frac{N+2}{N}\,
dx}{\sqrt{x^2+1}^3}=\frac{1}{\sqrt{\pi}}\Gamma\left(\frac{N-2}{2N}\right)
\Gamma\left(\frac{N+1}{N}\right)\ .
\]
For $N\gg 1$ we have $\alpha_N=1$. The flat-band result,
Eq.~(\ref{Delta-FB}), is recovered if the number of layers is
$N\gg 2\ln (\gamma_1/\Delta_0)$. The coherence length for a finite system
is $\xi_0\sim \hbar /p_\Delta$. It approaches $\hbar/p_{\rm FB}$
for $N\to \infty$.

\begin{figure}[h] 
\centering
\includegraphics[width=8cm]{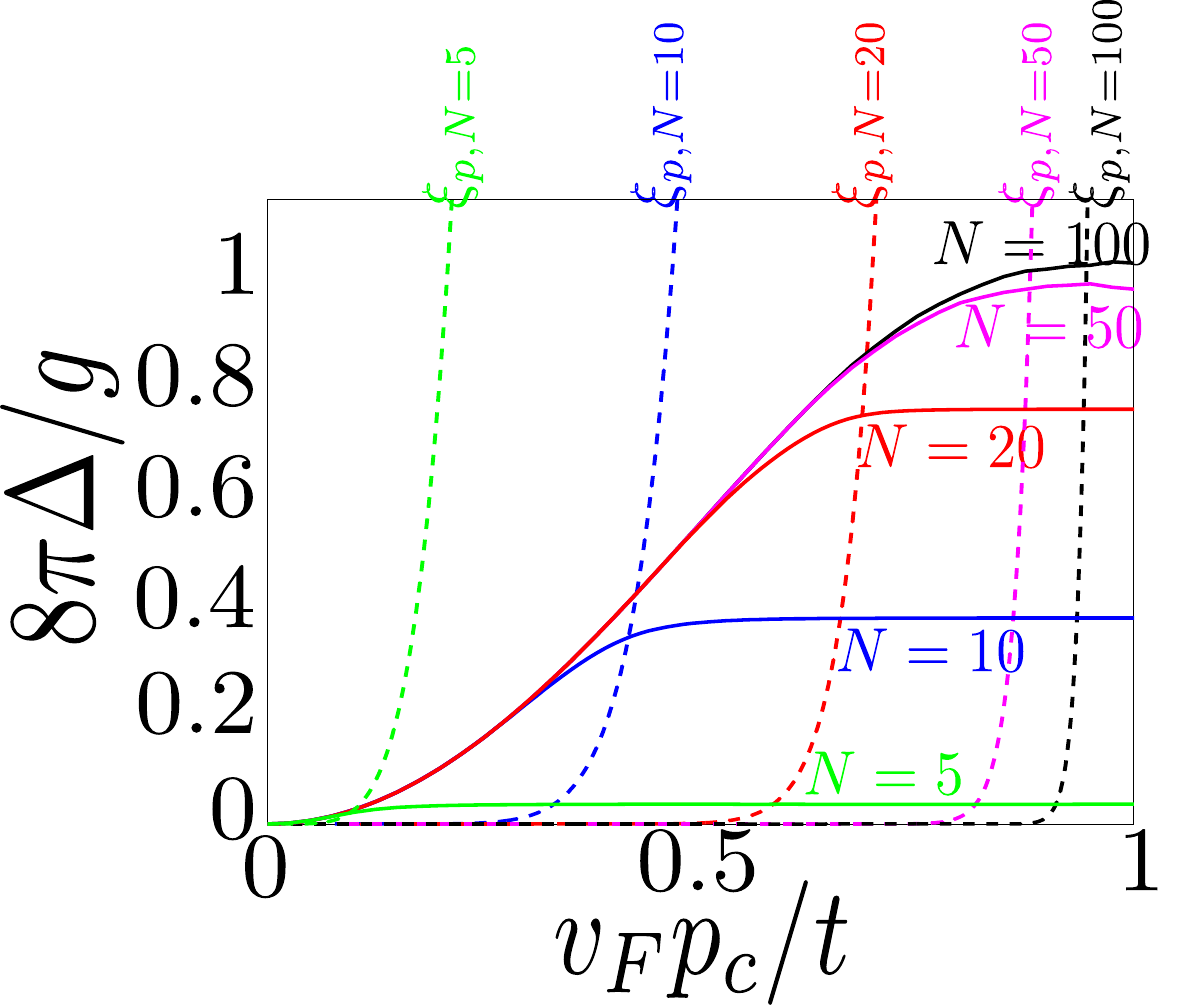}
 \caption{(Color online) Zero-temperature gap as a
 function of the momentum cutoff $p_c$ for various $N$ (solid
 lines). The gap saturates at $p_c\sim p_\Delta$ and approaches
 Eq.~(\protect\ref{Delta-FB}) for $N\to \infty$. The dashed lines
 show the dispersion $\xi_p$ for each $N$. Here $g=0.01\gamma_1$ and the higher-order couplings $\gamma_3$ and $\gamma_4$ have been set to zero.} \label{fig-gap}
 \end{figure}

The gap obtained by numerical integration of Eq.~(\ref{OP1})
with a cut-off $p_{co}$ is plotted in Fig.~\ref{fig-gap}. It shows how the gap is independent of the number of layers as long as $\Delta > \xi_{p_{c0}}$, and saturates for $\xi_{p_{c0}}> \Delta$. As the number of layers $N$ or the coupling energy $g$ increases, this threshold moves to larger values of cutoff momenta. 

\subsection{BCS-like surface superconductivity for a quadratic spectrum}
\label{subsec:quadratic}

In this section we consider a system with an infinite number of layers, such that $\xi_p=0$. The spectrum has a weak dispersion Eq. (\ref{eq:bdgspectrum}) due to a large effective mass $m^*$ determined by the inter-layer coupling $\gamma_4$.
In this case Eq. (\ref{Delta-gen}) yields for $T=0$
\begin{eqnarray}
1 &=&\frac{v^2  g}{\gamma_1^2} \int_{p<p_{FB}}\frac{d^2 p}{(2\pi)^2} \frac{\left(1-v^2p^2/\gamma_1^2 \right)^2 }{\sqrt{\mu_p^2+|\Delta|^2(1-v^2p^2/\gamma_1^2)^2}}\ .  
\label{eq:selfcons}
\end{eqnarray}

A simple qualitative expression can be written neglecting the normalization factor $1-v^2p^2/\gamma_1^2$:
\begin{eqnarray*}
1 &=&\frac{v^2  g}{\gamma_1^2} \int_{p<p_{FB}}\frac{d^2 p}{(2\pi)^2}  \frac{1}{\sqrt{\mu_p^2+|\Delta|^2}}\\
&=&\frac{g}{4\pi \alpha}\left[ {\rm Arsinh} \frac{\alpha -\mu}{\Delta} + {\rm Arsinh} \frac{ \mu}{\Delta}\right]\ ,
\end{eqnarray*}
where
$
\alpha=2\gamma_1(\gamma_4/\gamma_0).
$
For $\mu =0$ and $\mu =\alpha$ we find
\[
\Delta =\frac{\alpha}{\sinh(4\pi \alpha/g)} \ .
\]
This is Eq.~(\ref{eq:selfcons1}) discussed in Sec.~\ref{sec:introduction}. In the limit $g\gg \alpha$ we get the flat band result $\Delta =g/4\pi$ (without the extra $1/2$ because of the absence of the normalization factor $1-v^2p^2/\gamma_1^2$). Whereas this function captures the qualitative aspects of the self-consistent gap (exponential suppression below $g \lesssim 4 \pi \alpha$, and flat-band limit for $g \gtrsim 4 \pi \alpha$), the exact numerical value for the gap needs to be found from the full equation \eqref{eq:selfcons}.


\begin{figure}[h]
\centering
\includegraphics[width=8cm]{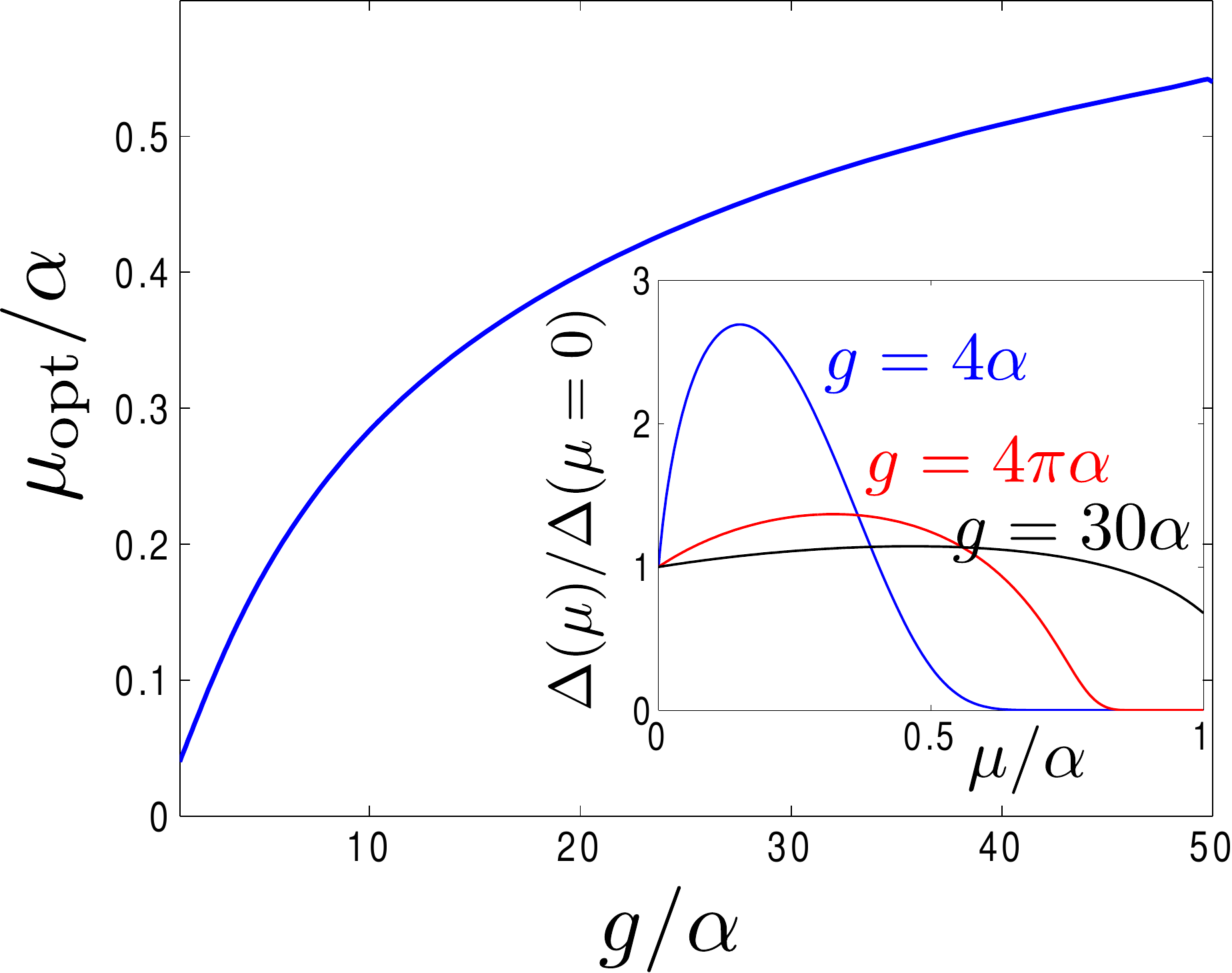}
\caption{Optimal doping level vs. the coupling constant $g$ for a given value of $\alpha$. Inset shows the dependence of the (normalized) self-consistent gap on the chemical potential for a few values of $g$. In the limit $g \gg 4\pi \alpha$, the scale for the $\mu$-dependence is $\Delta > \alpha$ rather than $\alpha$, but the optimum is still found around $\mu \approx 0.7\alpha$. }
\label{fig:optdoping}
\end{figure}

For a weak interaction $\Delta \ll \alpha$ the gap Eq.~(\ref{eq:selfcons1}) has a BCS-like form $\Delta =\alpha e^{-1/\lambda_2}$  where the coupling constant is
$
\lambda_2\sim g/\alpha
$. Using the estimate Eq. (\ref{g-estim}) for $g$  we find that this is a much larger coupling constant
\[
\lambda_2 \sim \lambda (\gamma_1/\gamma_4) \gg \lambda
\]
than what one would have for the usual bulk superconductivity. The coherence length has its usual form
\[
\xi_0  =\hbar v_g/\Delta  \sim \hbar p_{FB}/m^* \Delta  \sim a_0(\gamma_0/\gamma_1) e^{1/\lambda_2}\ ,
\]
which is much longer than the interatomic distance $a_0$.
\begin{figure}[h]
\centering
\includegraphics[width=8cm]{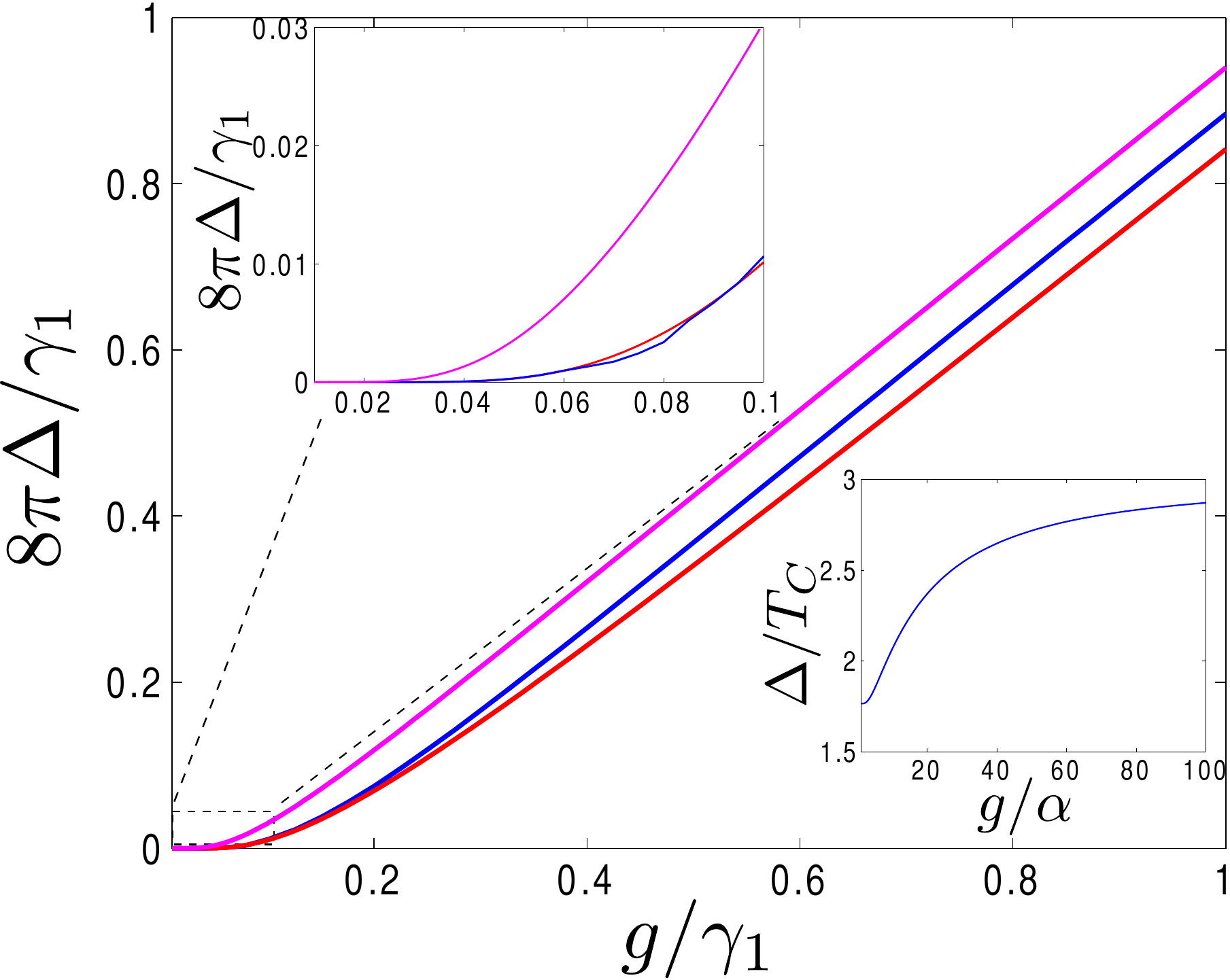}
\caption{Self-consistent surface gap function vs. coupling $g$. The blue line shows the results from the exact numerics in the case of $N=20$ layers and using $\mu=0$, the red line the numerical solution from Eq.~\eqref{eq:selfcons} at $\mu=0$ and the magenta line at $\mu=\mu_{\rm opt}$ plotted in Fig.~\ref{fig:optdoping}. For large coupling, the surface gap tends towards the flat-band limit $\Delta \propto g$. For coupling $g \lesssim 4\pi \alpha$, the gap becomes exponentially suppressed, $\Delta \propto \exp(-4\pi\alpha/g)$. The lower inset shows the ratio of $\Delta$ and the critical temperature $T_c$ as a function of the ratio $g/\alpha$ computed from Eq.~(\ref{Delta-gen}). For low coupling it equals to the regular BCS value 1.764, whereas in the flat band regime $g \gg 4 \pi \alpha$ it tends towards 3.}
\label{fig:selfconsdelta}
\end{figure}

Inserting a non-zero chemical potential $\mu$ in Eq.~\eqref{eq:selfcons} enables us to find an optimal doping with which the gap $\Delta$ is maximized. Such an optimal doping depends on the ratio $g/\alpha$, vanishing at $g\ll \alpha$ and saturating to a finite value $\mu_{\rm opt} \approx 0.7\alpha$ at $g \gg \alpha$. The optimal doping is plotted in Fig.~\ref{fig:optdoping}, along with the dependence of the gap on the chemical potential for a few values of $g$. This dependence makes the critical temperature sensitive to the presence of various impurities, complying with the reports of high-temperature superconductivity in doped graphite in \cite{Kopelevich01}.

\begin{figure}[h]
\centering
\includegraphics[width=8cm]{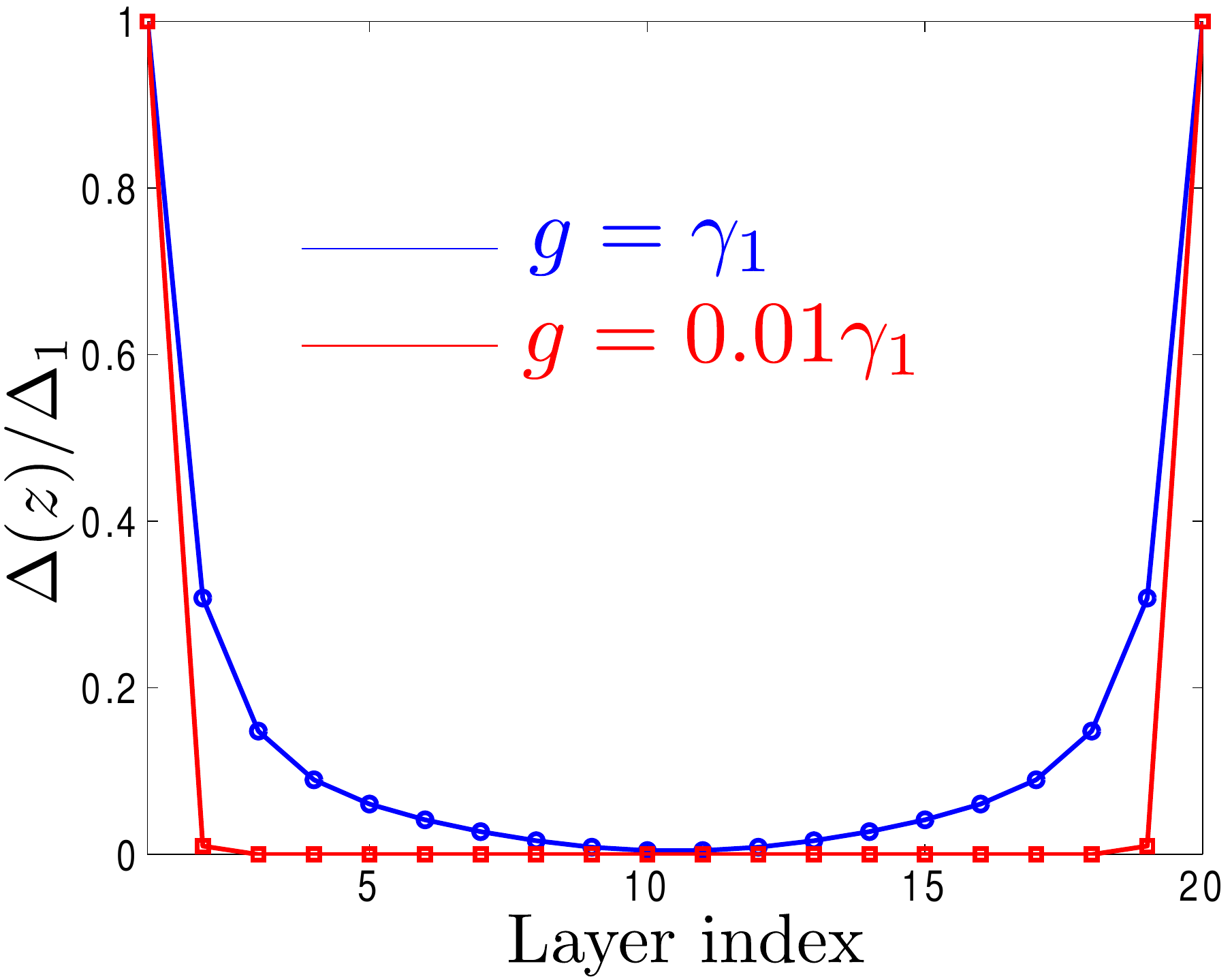}
\caption{Profile of the gap function $\Delta$ for two different coupling strengths for $N=20$ layers. The lines are guides to the eye, as $\Delta$ is only defined in the discrete points marked with circles ($g=\gamma_1$) and squares ($g=0.01 \gamma_1$). In the case $g=\gamma_1$ the induced gap is already so large that high-momentum states with $v_F p \sim \gamma_1$ contribute to superconductivity, and therefore superconductivity extends into the bulk of the material.}
\label{fig:Deltaprofile}
\end{figure}


These approximations are compared to the exact numerical solution of the self-consistency equation (\ref{OP1}), using the full Hamiltonian, Eq.~(\ref{BdGHamilt}) in Fig.~\ref{fig:selfconsdelta}. Moreover, Fig.~\ref{fig:Deltaprofile} shows the position dependence of the gap away from the surface state for two different coupling strengths. For low values of the coupling it extends into the bulk
only over a few interlayer distances due to a decay of the wave
functions. Taking this into account we have chosen the model below Eq.~(\ref{BdG-surf-a+}), in which the order parameter is
nonzero only on the outermost layers.

\subsection{Supercurrent}

Absence of dispersion in a flat band raises the questions of
superconducting velocity and of the supercurrent: Can they be
nonzero and, if they can, what is then the magnitude of the
critical current? In this section we address the problem of
supercurrent associated with the surface superconductivity in the
flat-band multilayered rhombohedral graphene. Based on the model
employed in \cite{KopninHeikkilaVolovik2011} for description of
the surface superconductivity the supercurrent was calculated in \cite{Kopnin2011} as a response to a small gradient of the order parameter phase $\Delta=|\Delta|e^{i{\mathbf k}{\mathbf r}}$ using an approach similar to that of \cite{KopninSonin10} for supercurrent in a single layer of graphene. The supercurrent appears to be finite; the critical current is
proportional to the superconducting zero-temperature gap, i.e., to
the critical temperature, and to the radius of the flat band in
the momentum space. Being produced by the surface
superconductivity, the total current through the sample is
independent of the sample thickness. Here we summarize the results obtained in \cite{Kopnin2011} for the flat band regime.

The current density along layer $n$ is
\begin{eqnarray}
{\mathbf j}_n&=&-e v_F\sum_{{\mathbf p},q}\left[ \hat u^\dagger
_n({\mathbf p}) \hat {\bm \sigma} \hat u_n({\mathbf p})+ \hat
v^\dagger _n({\mathbf p}) \hat {\bm \sigma} \hat
v_n({\mathbf p}) \right](1-2f_{{\mathbf p}}) \ . \quad \label{currentN}
\end{eqnarray}
Here $q$ labels different states for given ${\mathbf p}$,  while
$f_{{\mathbf p}}$ is the distribution function.
The current operator in Eq. (\ref{currentN}) couples the states at
different sublattices and thus contains the overlaps of the wave
functions localized at different surfaces.
Let us consider a large but finite number of layers. Equation (\ref{psi}) tells us that the wave function for each sublattice  decays away from the corresponding outer surface of the sample. As a result, the product of two wave functions at different sublattices in Eq. (\ref{currentN}) for the current, in addition to a decaying part, contains a contribution proportional to $(pv_F/\gamma_1)^N=\xi_p/\gamma_1$ that does not depend on the distance from the surface. As is seen from Eq. (\ref{DeltavsN}) the energy $\xi_p$ is of the order of the energy gap $\Delta$ at the surface. It is thus a characteristic energy associated with the superconducting coherence between the two surfaces. This coherence persists in the bulk even though the wave functions and the order parameter itself decay. This coherence produces a supercurrent that flows uniformly through the sample but with the density that is inversely proportional to the number of layers, see \cite{Kopnin2011}. The full current through the sample is
\begin{eqnarray*}
{\mathbf I} =\frac{2 e w \Delta \ln (\gamma_1/\Delta) {\mathbf k} }{\pi\hbar } \ ,
\end{eqnarray*}
where $w$ is the width of the sample. 
The full current does not depend on the sample thickness $Nd$ as
expected for surface superconductivity. The critical
current is determined by ${\rm max}(k) \sim \xi_0^{-1}$ where the
coherence length is $\xi _0 \sim
\hbar /p_{\rm FB}=\hbar v_{F}/\gamma_1$,
\[
I_c\sim \frac{e w\Delta \ln (\gamma_1/\Delta) p_{\rm FB}}{\hbar^2}\ .
\]

For nonzero $\mu$ we find in the same way as in
\cite{KopninSonin10}
\begin{eqnarray}
{\mathbf I}&=&\frac{ e w\ln (\gamma_1/\Delta) {\mathbf  k} }{\pi\hbar }\left[
\sqrt{|\mu|^2 +|\Delta|^2} \right. \nonumber \\
&&+\left. \frac{|\Delta|^2}{|\mu|}\ln\left(\frac{|\mu|+
\sqrt{|\mu|^2 +|\Delta|^2}}{|\Delta|}\right)\right]\ .
\label{curr-mu0}
\end{eqnarray}
This equation holds for $T\ll |\Delta|$. 

The supercurrent is thus finite despite the
absence of dispersion of the excitation spectrum. The critical
current is proportional to the zero-temperature gap, i.e., to the
superconducting critical temperature and to the size of the flat
band in the momentum space.

\subsection{Effect of fluctuations}

In our analysis above we employ the mean-field approximation. The quality of this approximation is determined by the Ginzburg number which is a measure of the relative magnitude of order-parameter fluctuations. For usual 3D superconductors the Ginzburg number is ${\rm Gi}\sim (T_c/E_F)^4 $; it is very small due to a small ratio of the critical temperature to characteristic energy of electrons which is the Fermi energy. 

In the case of flat band surface superconductivity, both the critical temperature and the characteristic energy of electrons are of the same order. Indeed, according to results of Sec.~\ref{subsec:flatband} the characteristic energy $\xi _p $ is of the order of $ \Delta \sim T_c$. As a result, the Ginzburg number is of the order of unity. The magnitude of the order-parameter fluctuation $\Delta_1$ can be estimated as follows. The free energy density of fluctuations in the flat-band regime is
\[
F_1 \sim \Delta_1 p_{FB}^2/\hbar^2\ .
\]
Since $\xi \sim \hbar/p_{FB}$ one has the total free energy of fluctuations ${\cal F}_1\sim \xi^2 F_1 \sim \Delta_1$. Comparing this energy with the thermal energy $T$ we find $\Delta_1 \sim T$. The thermal fluctuations are thus of the order of the mean-field gap $\Delta_1 \sim \Delta_0$ for $T\sim T_c$; they freeze out only for $T\ll T_c$. Therefore, the mean-filed approach is not exact for the flat-band regime. It works well, however, for the regime when the quadratic spectrum dominates.

If the superconductivity is dominated by the normal spectrum  with a quadratic dispersion, the fluctuation free energy density for $T$ not too close to $T_c$ is
\[
F_1 \sim \frac{\nu_2 \Delta_1^2}{2} = \Delta_1^2\frac{ \gamma_0\gamma_1}{16 \pi \hbar^2 v_F^2 \gamma_4}\ ,
\]
where $\nu_2 =m^*/2\pi \hbar^2$ is the 2D DOS and the effective mass $m^*$ is determined by Eq. (\ref{mass}).
If the coherence length is $\xi_0 =\hbar v_g \Delta_0^{-1}$ where $\Delta_0$ is the mean-field gap, the full energy in an area $\pi \xi_0^2$ is
\[
{\cal F}_1\sim \pi \xi_0^2 F_1 =
\frac{\Delta_1^2}{\Delta_0^2} \frac{v_g^2 \gamma_0\gamma_1}{16  v_F^2 \gamma_4}= \frac{\Delta_1^2}{\Delta_0^2} \frac{ \gamma_4\gamma_1}{\gamma_0}\ .
\]
Since $ {\cal F}_1\sim T$ we find
\[
\frac{\Delta_1^2}{\Delta_0^2}={\rm Gi} \sim  \frac{T \gamma_0}{ \gamma_4\gamma_1} \sim \frac{\Delta_0 \gamma_0}{ \gamma_4\gamma_1}\ .
\]
If the quadratic dispersion dominates, the order parameter is $\Delta_0 \ll \gamma_1\gamma_4/\gamma_0$, as follows from Eq.~(\ref{eq:selfcons1}). Therefore, the Ginzburg number ${\rm Gi}= e^{-1/\lambda_2}\ll 1$, and the average fluctuation of the order parameter is small compared to its mean-field value. However, at the crossover point to the flat band regime, $\Delta_0 \sim \gamma_1\gamma_4/\gamma_0$, and the fluctuation becomes of the same order as the mean-field value.

\subsection{Twinning boundary superconductivity}
The flat band states do not have to appear only at the surfaces of rhombohedral graphite. Similar flat bands can appear at twinning boundaries between different rhombohedrally stacked regions, or between a rhombohedral and Bernally stacked region\cite{HeikkilaKopninVolovik10}. The presence of such flat bands can be justified from the fact that at such twinning boundaries, the bulk topological charge $N_1({\mathbf p})$ discussed in \cite{HeikkilaVolovik10-1} changes sign or turns from a finite value to zero. Hence an asymptotically zero-energy state has to form at such boundaries for those momenta ${\mathbf p}$ in the $p_x,p_y$ plane for which $N_1$ changes. 

We do not discuss the microscopic features of such twinning boundary flat bands here, but present only a numerically calculated gap function for such a system in a few example cases: (a) a stacking fault in rhombohedral stacking, where at one point in the sample the new layer stacked on top of the rhombohedrally stacked layers has its B-atom on top of the A atom of the previous layer, which on the other hand was on top of the B-atom of the previous layer (onset of Bernal stacking, but continued by rhombohedral stacking), (b) a few such Bernally stacked layers on top of each other, and (c) rhombohedral stacking fault in an otherwise Bernally stacked graphite. The corresponding nearest-neighbour Hamiltonians around the stacking fault region are of the form
\begin{align*}
&H_a=\begin{pmatrix} 
H_{\rm rhg}(N_1,\sigma_+) & \hdotsfor{5}\\
 \gamma_1 \sigma_- & v_F \sigma \cdot {\mathbf p} & \gamma_1 \sigma_+ & \hdotsfor{3}\\
\hdotsfor{2} & \gamma_1 \sigma_- & v_F \sigma \cdot {\mathbf p} & \gamma_1 \sigma_- & \dots\\
\hdotsfor{3} & \gamma_1 \sigma_+ & v_F \sigma \cdot {\mathbf p} & \gamma_1 \sigma_- \\
\hdotsfor{5} & H_{\rm rhg}(N_2,\sigma_-) & 
\end{pmatrix}\\
&H_b=\begin{pmatrix} 
H_{\rm rhg}(N_1,\sigma_+) & \hdotsfor{5}\\
\gamma_1 \sigma_- & v_F \sigma \cdot {\mathbf p} & \gamma_1 \sigma_+ & \hdotsfor{3}\\
\dots & \gamma_1 \sigma_- & v_F \sigma \cdot {\mathbf p} & \gamma_1 \sigma_- & \hdotsfor{2}\\
\hdotsfor{2} & \gamma_1 \sigma_+ & v_F \sigma \cdot {\mathbf p} & \gamma_1 \sigma_+ & \hdotsfor{1}\\ 
 \hdotsfor{3} & \gamma_1 \sigma_- & v_F \sigma \cdot {\mathbf p} & \gamma_1 \sigma_- \\
 \hdotsfor{5} & H_{\rm rhg}(N_2,\sigma_+)  
\end{pmatrix}\\
&H_c=\begin{pmatrix}
v_F \sigma \cdot {\mathbf p} & \gamma_1 \sigma_+ & \hdotsfor{5}\\
\gamma_1 \sigma_- & v_F \sigma \cdot {\mathbf p} & \gamma_1 \sigma_- & \hdotsfor{4}\\
\dots & \gamma_1 \sigma_+ & v_F \sigma \cdot {\mathbf p} & \gamma_1 \sigma_+ & \hdotsfor{3}\\
\hdotsfor{3} & H_{\rm rhg}(N,\sigma_+) & \hdotsfor{3}\\
\hdotsfor{3} & \gamma_1 \sigma_+ & v_F \sigma \cdot {\mathbf p} & \gamma_1 \sigma_+ & \dots\\
\hdotsfor{4} & \gamma_1 \sigma_- & v_F \sigma \cdot {\mathbf p} & \gamma_1 \sigma_-\\
\hdotsfor{5} & \gamma_1 \sigma_+ & v_F \sigma \cdot {\mathbf p}
\end{pmatrix} \ .
 \end{align*}
Here the dots denote zeros in the matrix and the matrices $H_{\rm rhg}(N,\sigma)$ are the Hamiltonians for $N$ layers of rhombohedrally stacked graphite with the coupling matrix $\sigma$ on the upper diagonal.

\begin{figure}[h]
\centering
\includegraphics[width=\linewidth]{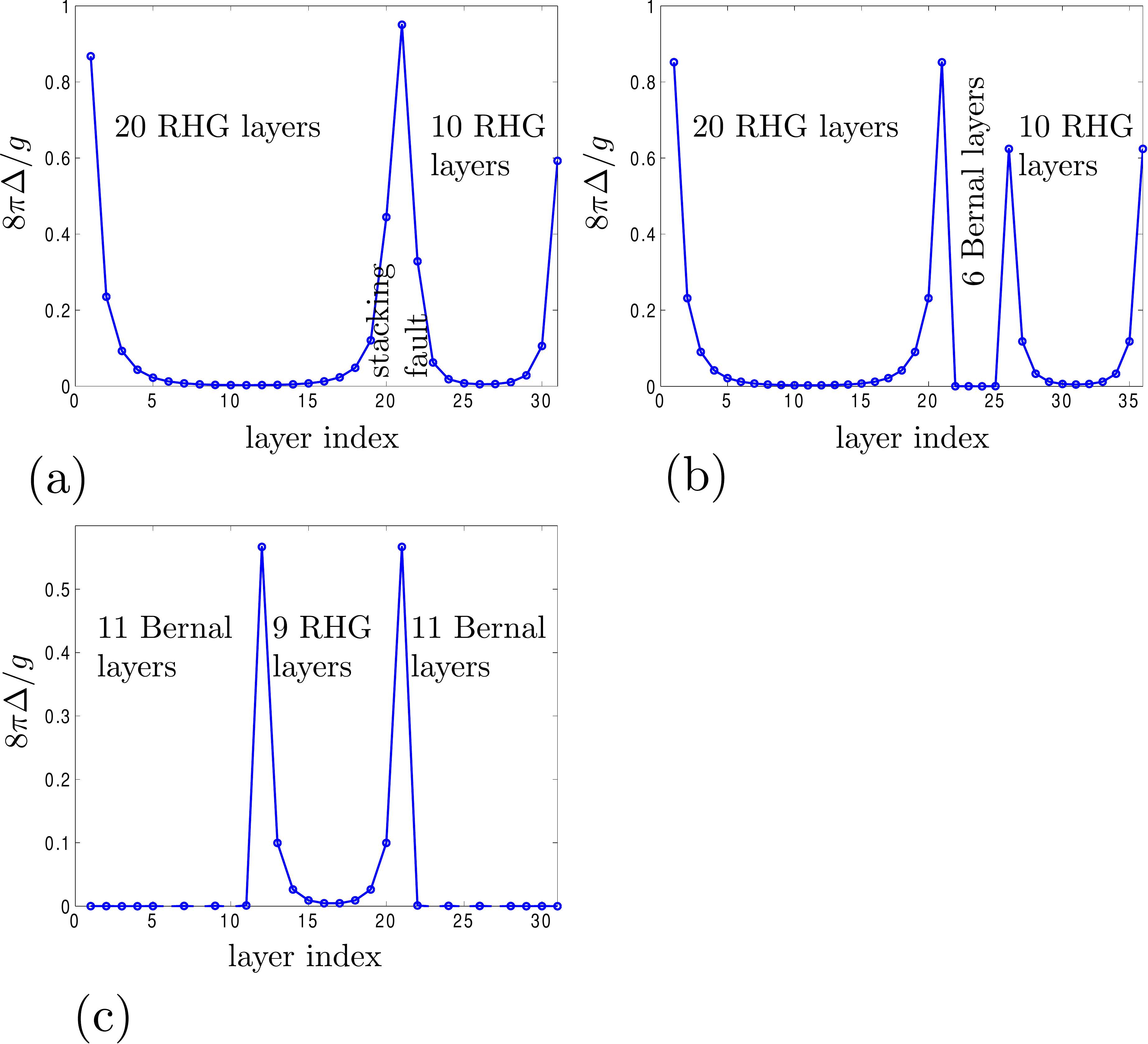}
\caption{Profile of the gap function $\Delta$ for example systems showing twinning boundaries: (a) Two rhombohedrally stacked graphite (RHG) multilayers coupled by a stacking fault. (b) Two rhombohedrally stacked graphite multilayers coupled by a Bernally stacked multilayer consisting of 6 layers. In both (a) and (b) the upper rhombohedrally stacked part contains 20 layers, and the lower part 10 layers. (c) Rhombohedrally stacked graphite multilayer sandwiched between two Bernally stacked multilayers. In all pictures, we have chosen $g=0.1 \gamma_1$ and disregarded the higher-order couplings $\gamma_3$ and $\gamma_4$.}
\label{fig:twindeltas}
\end{figure}

The corresponding profiles of the superconducting order parameter are plotted in Fig.~\ref{fig:twindeltas}. This picture clearly shows how interface superconductivity shows up at the twinning boundaries. 

\section{Summary}

The flat band with infinite DOS emerges in semi-metals with topologically protected nodal lines. This flat band promotes surface superconductivity with $T_c$ proportional to the pairing interaction strength and to the area of the flat band in the momentum space which is determined by the projection of the nodal
line onto the surface. Topologically protected flat
bands may also appear on interfaces, twin boundaries and grain
boundaries in bulk 3D topological materials leading to an enhanced
bulk $T_c$. 

Rhombohedral graphite is a promising candidate for a system with topologically protected surface flat band at the Fermi energy. Earlier\cite{HeikkilaVolovik10-1,HeikkilaKopninVolovik10,KopninHeikkilaVolovik2011} we have shown that, within the nearest-neighbour approximation, the rhombohedral graphite has a flat band for surface states, and these surface states support high-temperature superconductivity with the superconducting order parameter concentrated around the surfaces. The corresponding critical temperature is proportional to the pairing interaction strength and can be thus considerably higher than the usual exponentially small critical temperature in the bulk. This is in strong contrast to single-layer graphene, where the density of states vanishes at the Dirac point and therefore superconductivity quite generally requires strong doping \cite{KopninSonin10,Profeta12,CastroNeto05,KopninSonin08,Uchoa07,BlackSchDoniach07,Honerkamp08,EinenkelEfetov11}, (see also review \cite{Kotov10} and references therein). 

However, next-nearest neighbour hoppings which are present in real rhombohedral graphite can break the exact topological protection and, therefore, the flat-band mechanism of superconductivity at sufficiently low values of the coupling constant can be destroyed. Here we have studied the detailed effect of these higher-order interactions and demonstrated that instead of the flat-band scenario they can provide another mechanism of surface superconductivity which is of the BCS type but with a much larger coupling constant than the usual superconductivity in bulk. This large coupling constant comes from a large DOS associated with a heavy effective mass of surface quasiparticles that is clearly distinguishable on the background of the flat band which would exist without the higher-order interactions. For strong coupling, however, even this system crosses to the regime of flat band superconductivity.

Indications towards surface superconductivity have been seen in experiments on graphite in the form of a small Meissner effect and of a sharp drop in resistance \cite{Kopelevich01,Esquinazi08}. The enhanced superconducting density has been also reported on twin boundaries in Ba(Fe$_{1-x}$Co$_x$)$_2$As$_2$ \cite{Moler2010}. This year, these findings have been ratified and experimentalists have seen zero resistance in graphitic samples up to temperatures of 175 K \cite{Ballestar12} and furthermore indications of even room-temperature superconductivity in specially prepared graphite samples \cite{Esquinazi12}. These observations are compatible with surface or interface superconductivity described by our theory. However, they would require at least the presence of rhombohedrally stacked graphite regions embedded inside otherwise Bernally stacked regions of graphite. 

Besides the top and bottom surfaces of rhombohedral graphite, Bernally stacked graphite should have similar types of flat bands emerging on their lateral surfaces \cite{reviewCastroNeto09,Volovikslide}. It is possible that these states also support high-temperature surface superconductivity. 

 Our predictions provide a criterion for the parameters needed to obtain the highest critical temperature. They can be used for the search or for an artificial fabrication of layered and/or twinned systems with high- and even room-temperature superconductivity.
%


We thank G.\ Volovik for helpful comments and the collaboration that initiated this project. We also acknowledge fruitful discussions with F. Mauri, A. Harju and M. Ij\"as.
This work is supported in part by the Academy of Finland and its
COE program 2012--2016, by the European Research Council (Grant
No. 240362-Heattronics), and by the Program ``Quantum
Physics of Condensed Matter'' of the Russian Academy of Sciences.


\begin{thebibliography}{99}

\bibitem[{Khodel and Shaginyan(1990)}]{Khodel1990}
V.A. Khodel and V.R.  Shaginyan (1990),
JETP Lett. \textbf{51}, 553 (1990).

\bibitem[{Volovik(1991)}]{NewClass}
G.E. Volovik (1991),
JETP Lett. \textbf{53}, 222 (1991).


\bibitem[{Shaginyan {\it et al.}(2010)}]{Shaginyan2010}
V.R. Shaginyan, M.Ya. Amusia, A.Z. Msezane, K.G. Popov (2010),
 Phys. Rep. {\bf 492}, 31--109 (2010).

\bibitem[{Gulacsi(2010)}]{Gulacsi2010}
Z. Gulacsi, A. Kampf and D. Vollhardt (2010),
Phys. Rev. Lett. {\bf 105}, 266403 (2010).

\bibitem[{Heikkil{\"a} et al.(2011)}]{HeikkilaKopninVolovik10} T.T. Heikkil{\"a}, N.B. Kopnin, and G.E. Volovik (2011), Pis'ma ZhETF {\bf 94}, 252-258 (2011);
arXiv:1012.0905.

\bibitem[{Ryu and Hatsugai(2002)}]{Ryu2002}
S. Ryu and  Y. Hatsugai (2002),
Phys. Rev. Lett. {\bf 89}, 077002 (2002).

\bibitem[{Schnyder and Shinsei(2010)}]{SchnyderRyu2010}
A.P. Schnyder and Shinsei Ryu (2010),
 arXiv:1011.1438;
 
\bibitem[{Brydon et al.(2011)}]{Brydon2011}
P.M.R. Brydon, A.P. Schnyder, and C. Timm (2011),
 arXiv:1104.2257.

\bibitem[{Guinea et al.(2006)}]{GuineaCNPeres06}  F. Guinea, A.H. Castro Neto, and N.M.R.
Peres (2006), Phys. Rev. B {\bf 73}, 245426 (2006).

\bibitem[{Heikkil{\"a} and Volovik(2011)}]{HeikkilaVolovik10-1} T.T. Heikkil{\"a} and G.E. Volovik (2011),
JETP Lett. {\bf 93}, 59--65 (2011).

\bibitem[{Mak et al.(2010)}]{MakShanHeinz2010} Kin Fai Mak, Jie Shan, and T.F. Heinz (2010),
Phys. Rev. Lett. {\bf 104}, 176404 (2010).

\bibitem[{Dora et al.(2011)}]{Dora2011}
  B. Dora, J. Kailasvuori and R. Moessner (2011),
arXiv:1104.0416.

\bibitem[{Kopnin and Salomaa(1991)}]{KopninSalomaa1991}
N.B. Kopnin and M.M. Salomaa (1991),
Phys. Rev. B {\bf 44}, 9667--9677 (1991).

\bibitem[{Volovik (2011)}]{Volovik2011}
G.E. Volovik (2011),
 JETP Lett. {\bf 93}, 66--69 (2011).

\bibitem[{Kopnin et al.(2011)}]{KopninHeikkilaVolovik2011} N.B. Kopnin, T.T. Heikkil{\"a}, and G.E. Volovik (2011), Phys Rev. B  {\bf 83}, 220503 (2011).

\bibitem[{Ricardo et al.(2001)}]{Kopelevich01} R. Ricardo da Silva, J.H.S. Torres, and Y.
Kopelevich (2001), Phys. Rev. Lett. {\bf 87} 147001, (2001).

\bibitem[{Esquinazi et al.(2008)}]{Esquinazi08} P. Esquinazi, N. Garc{\'i}a, J.
Barzola-Quiquia, P. R{\"o}diger, K. Schindler, J.-L. Yao, and M.
Ziese (2008), Phys. Rev. B {\bf 78}, 134516 (2008);
\bibitem[{Dusari et al.(2010)}]{Dusari10} S. Dusari, J. Barzola-Quiquia and P. Esquinazi (2010),
arXiv:1005.5676.
%
\bibitem[{Scheike et al.(2012)}]{Esquinazi12} T. Scheike, W. B{\"o}hlmann, P. Esquinazi, J. Barzola-Quiquia, A. Ballestar, and A. Setzer (2012), Advanced Materials, {\bf 24}, pp. (2012).

\bibitem[{Ballestar et al.(2012)}]{Ballestar12} A. Ballestar, J. Barzola-Quiquia, and P. Esquinazi (2012), arXiv:1206.2463.

\bibitem[{Castro Neto et al.(2009)}]{reviewCastroNeto09} A. H. Castro Neto,
F. Guinea, N. M. R. Peres, K. S. Novoselov, and A. K. Geim (2009), Rev. Mod. Phys. {\bf 81}, 109 (2009).

\bibitem[{Dresselhaus and Dresselhaus(2002)}]{Dresselhaus02} M. S. Dresselhaus and G. Dresselhaus (2002), Advances in Physics {\bf 51}, 1 (2002). 

\bibitem{KopninIjasHarjuHeikkila12} N.B. Kopnin, M. Ij\"as, A. Harju, and T.T. Heikkil\"a, [arXiv:1210.7595].

\bibitem[{Arovas and Guinea(2008)}]{arovas08} D. Arovas and P. Guinea (2008), Phys. Rev. B {\bf 78}, 245416 (2008).

\bibitem[{McClure(1969)}]{McClure69} J.W. McClure (1969), Carbon {\bf 7}, 425 (1969).



\bibitem[{Kopnin(2011)}]{Kopnin2011} N.B. Kopnin (2011), Pis'ma ZhETF {\bf 94}, 81 (2011)[JETP Letters {\bf 94}, 81 (2011)].

\bibitem[{Kopnin and Sonin(2010)}]{KopninSonin10} N.B. Kopnin and E.B. Sonin (2010), Phys. Rev.
B {\bf 82}, 014516 (2010).

\bibitem[{Uchoa et al.(2005)}]{CastroNeto05} B. Uchoa, G.G. Cabrera, and A.H. Castro
Neto (2005), Phys. Rev. B, {\bf 71}, 184509 (2005).

\bibitem[{Kopnin and Sonin(2008)}]{KopninSonin08} N.B. Kopnin and E.B. Sonin (2008), Phys. Rev.
Lett. {\bf 100}, 246808 (2008).

\bibitem[{Profeta et al.(2012)}]{Profeta12} G. Profeta, M. Calandra and F. Mauri (2012), Nature Phys. {\bf 8}, 131 (2012).

\bibitem[{Uchoa and Cstro Neto(2007)}]{Uchoa07} B. Uchoa and A. H. Castro Neto (2007), Phys. Rev.
Lett. {\bf 98}, 146801 (2007); 

\bibitem[{Black-Schaffer and Doniach(2001)}]{BlackSchDoniach07}A. M. Black-Schaffer and S.
Doniach (2007), Phys. Rev. B, {\bf 75}, 134512 (2007); 

\bibitem[{Honerkamp(2008)}]{Honerkamp08} C. Honerkamp (2008),
Phys. Rev. Lett. {\bf 100}, 146404 (2008); 

\bibitem[{Einenkel and Efetov (2011)}]{EinenkelEfetov11} M.\ Einenkel and K.B.\ Efetov (2011), Phys. Rev. B {\bf 84}, 214508 (2011)


\bibitem[{Kotov et al.(2010)}]{Kotov10} V.N.
Kotov, B. Uchoa, V.M. Pereira, A.H. Castro Neto, and F. Guinea (2010),
arXiv: 1012.3484, and references therein.

\bibitem[{Kalisky et al.(2010)}]{Moler2010}  B. Kalisky, J.R. Kirtley, J.G. Analytis, Jiun-Haw Chu, A. Vailionis, I.R.
Fisher, K.A. Moler (2010),
Phys. Rev. B {\bf 81}, 184513 (2010).

\bibitem[{Volovik (2011)}]{Volovikslide} See slide 32 of http://www.pdmi.ras.ru/EIMI/2011/STMP/presentations/Volovik.pdf.









\end{thebibliography}
\end{document}